\documentclass[11pt,a4paper, final, twoside]{article}
\usepackage{amsmath}
\usepackage{longtable}
\usepackage{fancyhdr}
\usepackage{amsthm}
\usepackage{amsfonts}
\usepackage{amssymb}
\usepackage{amscd}
\usepackage{latexsym}
\usepackage{amsthm}
\usepackage{graphicx}
\usepackage{graphics}
\usepackage{afterpage}
\usepackage[colorlinks=true, urlcolor=blue,  linkcolor=black, citecolor=black]{hyperref}
\usepackage{color}
\usepackage{pgf, pgfarrows, pgfnodes, pgfbaselayers}
\usepackage{tikz}
\usetikzlibrary{shapes}
\usetikzlibrary{arrows}

\newcommand{\HRule}{\rule{\linewidth}{1pt}}

\theoremstyle{proposition}

\theoremstyle{definition}

\theoremstyle{remark}

\numberwithin{equation}{section}
\begin{document}
\hyphenpenalty=100000


\begin{flushright}

{\Large \textbf{\\Mathematical Model of Colorectal Cancer with Monoclonal Antibody Treatments }}\\[5mm]
{\large \textbf{L.G. dePillis$^\mathrm{*1}$\footnote{\emph{*Corresponding author: E-mail: depillis@hmc.edu}},
H. Savage$^\mathrm{2}$\\[1mm]
and A.E. Radunskaya$^\mathrm{3}$}}\\[1mm]
$^\mathrm{1}${\footnotesize \it Dept. of Mathematics,\\ Harvey Mudd College \\ Claremont, Califoria, USA}\\ 
$^\mathrm{2}${\footnotesize \it Dept. of Mathematics,\\ Harvey Mudd College \\ Claremont, Califoria, USA}\\ 
$^\mathrm{3}${\footnotesize \it Dept. of Mathematics,\\ Pomona College \\ Claremont, Califoria, USA}
\end{flushright}

\begin{flushleft}\fbox{%
\begin{minipage}{1.3in}
{\slshape \textbf{Research Article}\/}
\end{minipage}}
\end{flushleft}

\begin{flushright}\footnotesize \it Received: 10 December 2013\\ 
Accepted: XX December 20XX\\
Online Ready: XX December 20XX
\end{flushright}
\HRule\\[3mm]

\newpage
{\Large \textbf{Abstract}}\\[4mm]
\fbox{%
\begin{minipage}{5.4in}{\footnotesize 

We present a new mathematical model of colorectal cancer growth and its response to
monoclonal-antibody (mAb) therapy.  Although promising, most mAb drugs are still in trial phases,
and the possible variations in the dosing schedules of those currently approved for use have
not yet been thoroughly explored.  To investigate the effectiveness of current mAb treatment schedules,
and to test hypothetical treatment strategies, we have created a system of nonlinear
ordinary differential equations (ODE) to model colorectal cancer growth and treatment.
The model includes tumor cells, elements of the host's immune response, and treatments.
Model treatments include the chemotherapy agent irinotecan and one of two monoclonal antibodies -
cetuximab, which is FDA-approved for colorectal cancer, and panitumumab, which is still
being evaluated in clinical trials.  The model incorporates patient-specific parameters to account for
individual variations in immune system strength and in medication efficacy against the tumor.
We have simulated outcomes for groups of virtual patients on treatment protocols for which
clinical trial data are available,
using a range of biologically reasonable patient-specific parameter values.
Our results closely match clinical trial results for these protocols. We also
simulated experimental dosing schedules, and have found new schedules which, in our simulations,
reduce tumor size more effectively than current treatment
schedules. Additionally, we examined the system's equilibria and sensitivity to parameter values.
In the absence of treatment, tumor evolution is most affected by the intrinsic tumor growth rate and carrying
capacity. When treatment is introduced, tumor growth is most affected by drug-specific PK/PD parameters.
} \end{minipage}}\\[1mm]
\footnotesize{\it{Keywords:} }\\[1mm] 


\afterpage{
\fancyhead{} \fancyfoot{} 
\fancyfoot[R]{\footnotesize\thepage}
\fancyhead[R]{\scriptsize\it British Journal of Medicine and Medical Research
{{X(X), XX--XX}},~2013 \\
 }}


\section{Introduction}\label{I1}


According to the American Cancer Society, 
colorectal cancer is the third most commonly
diagnosed cancer and the third leading cause of cancer death in both women and men in the United States~\cite{cancer.org_2013}. 
Monoclonal antibodies have been explored as an adjuvant treatment for colorectal cancer, but there are still many unanswered questions about their effectiveness and optimal use. The goal of this work is to contribute to the understanding of how best to incorporate monoclonal antibodies into colorectal cancer treatment. We present a system of nonlinear ordinary differential equations (ODEs),
\footnote{
Acronyms: {mAb}: {monoclonal antibody}, 
{ODE}: {ordinary differential equation},
{NK}:{Natural Killer cell},
{CD8$^+$}:{cytotoxic T-cell},
{CTL}:{cytotoxic T-lymphocyte (often equivalent to CD8$^+$ T-cell)},
{ADCC}:{antibody-dependent cellular cytotoxicity},
{EGF}:{endothelial growth factor},
{EGFR}:{endothelial growth factor receptor},
{CDC}:{complement-dependent cytotoxicity}
}
that models the growth of a colorectal tumor, its interactions with the host's immune system, and the effects of three treatment options: the chemotherapy drug irinotecan, and two monoclonal-antibody (mAb) treatments, cetuximab and panitumumab. We use this model to run clinical trial simulations over cohorts of virtual 
patients with varying response rates. After validating our outcomes against published clinical trial data, we then explore alternate
hypothetical treatment scenarios.

\subsection*{Colorectal Cancer}
Monoclonal antibody therapies, which are targeted cancer therapies, are being tested in clinical trials to address colorectal tumors that
are chemotherapy-refractory.
	Monoclonal antibodies are small antibodies that are manufactured to bind to specific proteins. Multiple protein targets can be used, but epithelial growth factor receptor (EGFR) is a common and useful choice. EGFRs are found in cell membranes in cells all over our body, and circulating epithelial growth factor (EGF) binds to this receptor and signals a cascade in the cell, resulting in cell proliferation. The increased growth rate in tumor cells is usually caused by multiple mutations, but a common mutation upregulates the number of EGFRs~\cite{gravalos_et_al_2009,siena_et_al_2009,martinelli_et_al_2009,cancer_tx_book}. 
Monoclonal antibodies targeting EGFRs are considered promising since many cancerous cells have the EGFR-upregulating mutation.

\begin{figure}[!ht]
\begin{center}
\includegraphics[height=.4\textheight]{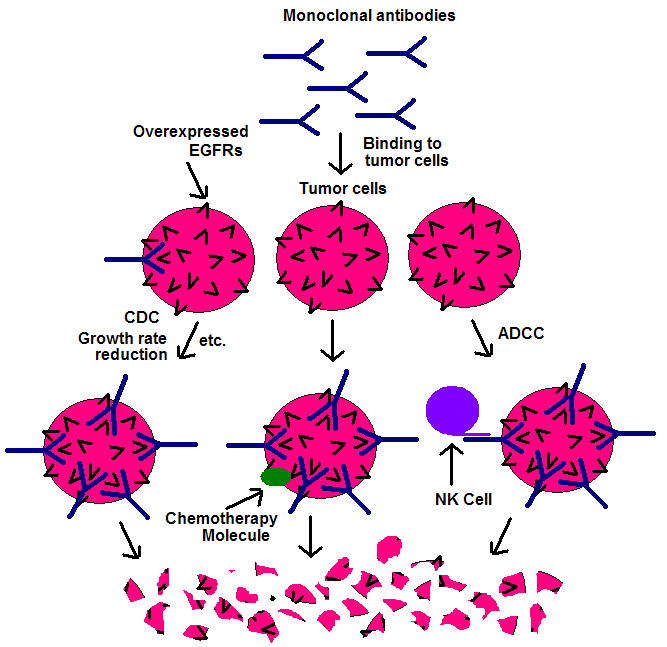}
\end{center}
\caption{{\bf Three methods of mAb-induced tumor cell death are represented in this model.} If an NK cell is present then the cell can undergo ADCC, if a chemotherapy molecule is present then the cell will increase death from the chemotherapy drug, and otherwise, the mAb molecule will cause tumor cell death on its own, through a variety of mechanisms.}
\label{fig:mab_use}
\end{figure}

	Cetuximab and panitumumab, both monoclonal antibodies that bind to EGFR and block EGF from binding, are two monoclonal antibodies 
that have been shown to have some degree of effectiveness
against colorectal cancer.
Cetuximab, used with or without the chemotherapy 
drug irinotecan, has been shown to improve survival times and quality of life~\cite{martinelli_et_al_2009}. It is an IgG1 antibody, a subclass of antibodies that is able to elicit antibody-dependent cellular cytotoxicity ({ADCC}) from Natural Killer ({NK}) cells, thus increasing the NK cells' cytotoxicity~\cite{martinelli_et_al_2009}. Panitumumab is a newer drug and has undergone fewer clinical trials. It has been shown to decrease tumor growth rate, but the clinical trials have not yet been able to confirm that it increases overall survival time~\cite{martinelli_et_al_2009}. Both cetuximab and panitumumab are able to activate the cascade known as complement dependent cytotoxicity ({CDC}), and both 
increase chemotherapy's toxicity to tumor cells by hindering their ability to reproduce~\cite{cancer_tx_book}. 
There are three main pathways for mAb induced tumor death (see Figure~\ref{fig:mab_use}): interactions between mAbs, NK cells, and tumor cells; 
interactions between mAbs, chemotherapy and tumor cells; 
and interactions only between mAbs and tumor cells, resulting in growth rate reduction, complement activation, 
and possibly other mechanisms for tumor death.

	Currently, monoclonal-antibody treatments 
are mainly used in patients with metastatic cancer, particularly when no other treatment has worked~\cite{martinelli_et_al_2009}. However, it is possible that with positive results from current clinical trials, monoclonal antibodies may become a more significant part of colorectal-cancer treatment.
The model presented here can shed further light on monoclonal antibody treatments by simulating clinical trials.

\subsection*{Previous Models}



A variety of approaches has been taken to the mathematical modeling of colorectal cancer growth and treatment. These include ODE models, spatial models,
and statistical models. A nice review  can be found in \cite{Ballesta2014}, where mechanistic models and phenomenological cell population models are discussed.  The primary focus of this review is to explore published models that include chemotherapy, with the end goal of optimizing therapy regimens.  In particular, in \cite{Ballesta2012} a  PK-PD model of irinotecan (CPT11) is combined with a compartment model to describe a whole-body physiologically based model for colon cancer in mice.  In the model presented in this paper, we include immunotherapy in addition to chemotherapy, using a phenomenological cell population model.

In \cite{johnston_et_al_2007}, Johnston {\em et al.} considered two different approaches to the modeling of cells in a colonic crypt. 
They first consider an age-structured model
which tracks the locations, properties, and ages of stem cells, transit cells, which move up the wall of the crypt to the surface, and differentiated cells. 
The resulting model was then simplified using a continuous approximation.
They found that the resulting ODE system provided a good approximation for the growth rate produced by the age-structured model for a sufficiently large time scale.
This work
which shows that an ODE system can be used to represent a 3D structure in the colon, motivated our choice of model for a colorectal tumor. 
However, the use of ODEs requires the simplifying assumption that the tumor is spatially homogeneous, and only tracks
tumor population changes over time. 
Since the measure of overall tumor size is used to indicate the strength of a patient's response to treatment in the clinic, it is reasonable
use tumor size as a measure of treatment efficacy in our model as well.

Other mathematical models of colonic cancer focus on the initiation of the disease.  For example, in \cite{komarova_2002} a mathematical model is developed that supports the hypothesis that two types of genetic instability can lead to the tumorigenesis in individuals with colorectal cancer.  More recently, Lo et al (\cite{lo_2013}) propose a mathematical model of the initiation of colorectal cancer that explores a possible link with colitis.

	The model presented here is an extension of the work of de Pillis {\em et al.}~\cite{de_pillis_et_al_2009}, in which a tumor-cell population, immune-cell populations, and drug concentrations are modeled with a system of nonlinear ODEs. The model by de Pillis {\em et al.} also includes patient-specific parameters representing the strength of the patient's immune system, and has been validated with published studies on mice and humans~\cite{de_pillis_et_al_2005}. It has successfully demonstrated the need for immunotherapy in addition to chemotherapy to prevent the tumor from growing again after drug therapies have been completed, and was used to study the importance of the patient-specific parameters in the effectiveness of immunotherapy treatment~\cite{de_pillis_et_al_2009,de_pillis_et_al_2008}. 
The new model 
includes terms for monoclonal-antibody treatment and its effects on the cell populations, and parameter values have been
adjusted to reflect dynamics specific to colorectal cancer.

\section{Mathematical Model}\label{I3}

The goal of this mathematical model is to describe tumor growth, immune response, and treatments, including chemotherapy and monoclonal antibody (mAb) treatments.
Our model tracks the following populations and quantities:
\begin{itemize}
\item Cell Populations
  \begin{description}
  \item[$T(t)$] the total tumor cell population;
  \item[$N(t)$] the concentration of NK cells per liter of blood (cells/L);
  \item[$L(t)$] the concentration of cytotoxic T lymphocytes ({CD$8^+$}) per liter of
    blood (cells/L);
  \item[$C(t)$] the concentration of lymphocytes per liter of blood,
    not in\-clud\-ing NK cells and active CD8$^+$T cells (cells/L).
  \end{description}
\item Medications and Cytokines
  \begin{description}
  \item[$M(t)$] the concentration of chemotherapy per liter of blood (mg/L);
  \item[$I(t)$] the concentration of interleukin per liter of blood
    (IU/liter);
  \item[$A(t)$] the concentration of monoclonal antibodies per liter
    of blood (mg/liter);
  \end{description}
The specific treatments that we will explore are the chemotherapeutic drug irinotecan (CPT11), 
and mAbs Cetuximab or Panatumumab.  
\item Treatments:
        \begin{description}
  \item[$v_M(t)$] the amount of irinotecan injected per day per
    liter of blood (mg/liter per day);
  \item[$v_A(t)$] the amount of monoclonal antibodies injected per
    day per liter of blood (mg/liter per day).
  \end{description}
\end{itemize}
In the following section, we give
a description of the equations describing the evolution of each of population. 
In Section \ref{J}  examples of the  evolution of  simulated cell populations over time are presented, and treatments and clinical trials were simulated. Finally, we present a parameter sensitivity analysis and discuss the results.  Details of the parameter estimation, a discussion of equilibria and their stability and further sensitivity analyses are given in the Supplementary Materials. 

\subsection*{Equations}\label{Equations}

%

The full system of ODEs of the model is given below.  The equations are based on the model proposed in \cite{de_pillis_et_al_2009}, with additions necessary to describe mAb and combination treatments.  These additional terms are shown in bold face. A summary of the purpose of each model term can be found in 
Tables~\ref{tab:tumor}-\ref{tab:meds}.

\begin{align}
  \frac{dT}{dt} & = & &  aT(1-bT) - (c+\boldsymbol\xi\mathbf{\frac{A}{h_1+A}})NT-DT\notag\\
  & & & {}- (K_T+\mathbf{K_{AT}A})(1-e^{ -\delta_TM})T-\boldsymbol\psi\mathbf{AT}\label{eq:tumor}\\
  \frac{dN}{dt} & = & &  eC-fN - (p+\mathbf{p_A\frac{A}{h_1+A}})NT +\frac{p_NNI}{g_N+I}\notag\\
  & & & {}- K_N(1-e^{-\delta_NM})N \label{eq:natural_killers}\\
  \frac{dL}{dt} & = & &  \frac{\theta mL}{\theta+I} + j\frac{T}{k+T}L-qLT+(r_1N+r_2C)T-\frac{uL^2CI}{\kappa+I}\notag\\
  & & & {}-K_L(1-e^{-\delta_LM})L +\frac{p_ILI}{g_I+I}\label{eq:killer_ts}\\
  \frac{dC}{dt} & = & &  \alpha- \beta C - K_C(1-e^{-\delta_CM})C \label{eq:lymphocytes}\\
  \frac{dM}{dt} & = & &  -\gamma M + v_M(t) \label{eq:chemo}\\
  \frac{dI}{dt} & = & &  -\mu_II+\phi C + \frac{\omega LI}{\zeta+I} \label{eq:interleukin}\\
  \mathbf{\frac{dA}{dt}} & = & &  -\eta\mathbf{A-}\lambda\mathbf{T\frac{A}{h_2+A} + v_A(t)}\label{eq:mabs}\\
  \intertext{where} 
 	D & = & &  d\frac{(L/T)^l}{s+(L/T)^l}.\label{eq:patient}
\end{align}%

\subsubsection*{Model Terms Describing Growth and Interactions}

Each of Equations \eqref{eq:tumor} - \eqref{eq:mabs} describes the time evolution of one of the eight system variables.  Each equation contains a growth, or source, term, and a decay term.  Most of the equations also contain interaction terms that describe how one population of cells or molecules affects another.  For example, in Equation \eqref{eq:tumor}, the tumor is assumed to grow logistically in the absence  of other cells or antibodies.  The competition term between tumor and NK cells follows a mass action law, where the effectiveness of the NK cells in killing tumor cells, or the {\it per cell} kill rate, is enhanced by the presence of monoclonal antibodies (see also the discussion below).  The interaction between cytotoxic T lymphocytes ({CTL}s) and tumor cells is described by a {\it ratio-dependent} law, articulated in Equation \eqref{eq:patient}.  The derivation of this term is described in detail in \cite{de_pillis_et_al_2005}.

A recruitment term is included for the tumor-specific CD$8^+$ T cells, as well as a production term in Equation \eqref{eq:interleukin} that reflects the increased presence of IL-2 when CTLs are present.
Interleukin, whose concentration is denoted by the variable $I(t)$, activates the production of 
NK cells and CD$8^+$ cells, indicated by the positive saturating terms in 
Equations \eqref{eq:natural_killers} and \eqref{eq:killer_ts}. 
However, IL-2 can also
also aid in the inactivation of CD$8^+$ cells. 
From Abbas et al. \cite{abbas_et_al_2005}, we find that the deactivation of CD8+T cells occurs 
through a pathway that requires IL-2 and the action of CD4+T cells 
(found in the circulating lymphocytes). Moreover, it occurs only at
high concentrations of activated CD8+T cells.
This deactivation is represented by the
term $-\frac{uL^2CI}{\kappa+I}$ in Equation \eqref{eq:killer_ts}.   

Also described in the model are the effects of a cytotoxic drug such as irinotecan.  This drug is assumed to have a detrimental effect on all of the cell populations.  For more details on the derivation of these terms see \cite{de_pillis_et_al_2009} and \cite{de_pillis_et_al_2005b}, and for parameter values, sources, and derivations see the Supplementary Materials.

\subsubsection*{Discussion of Terms Describing Treatment}\label{TreatmentTerms}

In this section we give details on terms in the model that were added to the one proposed in \cite{de_pillis_et_al_2009}.
	In Equation~\eqref{eq:tumor}, three terms represent the three pathways of mAb-induced tumor-cell death (see Figure~\ref{fig:mab_use}). 
\begin{itemize}
\item The term $-\xi\frac{A}{h_1+A}NT$ represents the rate of tumor-cell death caused by ADCC. Some monoclonal antibodies have protein structures which, when bound to a tumor cell, allow them to simultaneously activate NK cells and to direct them to the invader~\cite{cancer_tx_book}. Thus, when a mAb/tumor-cell complex and NK cell meet, the tumor cell is more likely to be killed than when an NK cell meets an unbound tumor cell. Kurai and colleagues~\cite{kurai_et_al_2007} found that cetuximab has a threshold concentration above which ADCC activity no longer increases. So, we assume that ADCC activity increases with mAb concentration until it becomes saturated, and we model this with a sigmoid function. 
\item The term $-K_{AT}A (1-e^{-\delta_TM})T$ represents the rate of chemotherapy-induced death of tumor cells, assisted by monoclonal antibodies. When tumor cells are not able to proliferate, they are much more susceptible to chemotherapy-induced death~\cite{cancer_tx_book}. So, when mAbs are bound to tumor cells, blocking their EGFRs and thus inhibiting tumor cell proliferation, they increase the tumor-cell death caused by chemotherapy. 
\item The term$-\psi AT$ accounts for the rate of tumor-cell death caused directly by tumor cell interactions with mAbs. This term includes tumor-cell death from CDC, from a reduction in EGF binding and thus tumor-growth rate, ~\cite{cancer_tx_book}.
\end{itemize}
	
The term $-p_A\frac{A}{h_1+A}NT$ in Equation~\eqref{eq:natural_killers} represents the rate of NK-cell death due to ADCC interactions with tumor cells and monoclonal antibodies. We assume that ADCC activity increases with mAb concentration until it becomes saturated. As with the term $-pNT$, it is assumed that NK cells experience exhaustion of tumor-killing resources after multiple interactions with tumor cells~\cite{bhat_watzl_2007}.	
	We note that we chose not to incorporate mAb interactions in Equations~\eqref{eq:killer_ts}, \eqref{eq:lymphocytes} and \eqref{eq:chemo}, since the literature suggests that
effects of mAbs are specific to tumor cells~\cite{rodriguez_et_al_2009,martinelli_et_al_2009,gravalos_et_al_2009,siena_et_al_2009,cancer_tx_book}.


	
The evolution of the monoclonal antibody population is described in Equation~\eqref{eq:mabs}.
The term $v_A(t)$ represents mAb treatments. Because mAbs are not produced naturally in the body, no additional growth terms are included. The term $-\eta A$ represents the natural degradation of the mAb protein in the body. The term $-\lambda T\frac{A}{h_2+A}$ represents the loss of available mAbs as they bind to tumor cells. 
mAbs have a very strong binding affinity for their target growth-factor receptors, and there are many growth factor receptors on every cell, so we assume that many mAbs are lost with each tumor cell. Also, we assume that the growth factor receptors are fully saturated when the mAb concentration is significantly higher than the growth factor receptor concentration. That is, we can approximate the number of mAbs lost with each tumor cell as the number of growth-factor receptors 
on that cell, as long as mAb concentration is not close to zero.
	

\section{Results}\label{J}

\subsection*{Clinical Trial Simulations for Common Treatment Regimens}

We used the model to explore expected responses to treatment at a population level.  In particular, we simulated response to treatment for patients with a range of immune `strength'.  
The effectiveness of the CD$8^+$ T-cells is described in the model by the term $D$ described in Equation \ref{eq:patient}.  In order to describe a group of patients with different immune strengths, we allow the three parameters in Equation \ref{eq:patient} to take on one of four values taken from a biologically reasonable range, ~\cite{de_pillis_et_al_2009}.  These three patient-specific  parameters are $d$, the maximum kill-rate by effector cells; $s$, the steepness of the effector-cell response to the presence of tumor; and $l$, a measure of the non-linearity of the response.  Table \ref{tab:tumor} lists the specific values used.  Using four different values for each of the three parameters yields 64 virtual patient types, each with a different immune system.  

To account for variation between patients in tumor response to therapy, we also varied the values of the parameters $K_T$, the rate of tumor-cell death from chemotherapy, and $\psi$, the rate of cell death induced by mAb agents.  For each simulation, the values of these parameters were randomly sampled from a distribution given by the density function 
$$p(x) = \frac{1}{3 x_{max}}(1-x/x_{max})^{-2/3}, 0 \leq x < x_{max},$$ 
where $x_{max}$ is the maximum value of each parameter, either $K_{max}$ or $\psi_{max}.$ (See Table \ref{tab:tumor}.)

In these clinical trial simulations, we assume that the 
simulated patients have slightly compromised immune systems after already having
been through other immuno-depleting therapies.   MAb therapy is currently used mainly as a last resort, after other treatments have been attempted unsuccessfully, so we expect the tumor population size to initially be large. Initial values for the state variables reflect this, with with a large initial tumor size, $T(0) = 10^9$ cells,  and relatively low levels of NK and CD$8^+$ lymphocytes.  
All initial values are given in Appendix \ref{S1}.
Simulated treatments were administered to each patient, represented by $v_M(t)$ and $v_A(t)$ in model equations \eqref{eq:chemo} and \eqref{eq:mabs}. 

Clinical trial simulations were run over the set of 64 virtual patients multiple times. 
Final tumor size and lymphocyte counts were recorded for each patient. 
Lymphocyte count was used as a marker for patient health---if the lymphocyte count dropped low enough for the patient to be 
considered grade 4 leukopenic,
the treatment was considered to be too harsh and not useful.   
This minimum lymphocyte count was determined to be $1.4\times10^8$,  based on the WHO criteria of grade 4 leukopenia being less than $10^9$ total white blood cells 
per liter, (\cite{welink_et_al_1999}, and see also the discussion of the parameter $K_C$ in the Supplementary Materials).
Final simulated tumor sizes  were categorized as a ``Complete Response'' (CR), ``Partial Response'' (PR), or ``No Response'' (NR). 
Tumors that continue to grow are categorized as NR, and any tumor smaller than $\approx 2.2$ mm in diameter is categorized as  CR. This value was chosen since it is 
significantly below the clinical detection level of 5 mm in diameter, \cite{recist}.
In our analysis, we assume a spherical, homogeneous tumor, so that $2.2$ mm in diameter corresponds to $2 \time 10^7$ cells. 
Finally, 
 those tumors that don't continue to grow, but are larger than $2\times 10^7$ cells, are categorized as PR. 
We compare the results of the simulated trials to those reported in  \cite{cancer_tx_book, lenz_2007, cunningham_et_al_2004, grothey_2006, gravalos_et_al_2009}.
 See Table \ref{tab:response_rates_common} for a summary of these clinical trial outcomes.
Note that the published clinical trial results for cetuximab and panitumumab that we used for comparison reported results as ``Response'' or ``No Response'' almost exclusively, so for our clinical trial simulations of the commonly used treatments, we group PR and CR together under ``Response''. 
This facilitates comparison between our simulation results and the results of reported clinical trials. 

\begin{figure}[!ht]
\begin{center}
\includegraphics[height=.2\textheight]{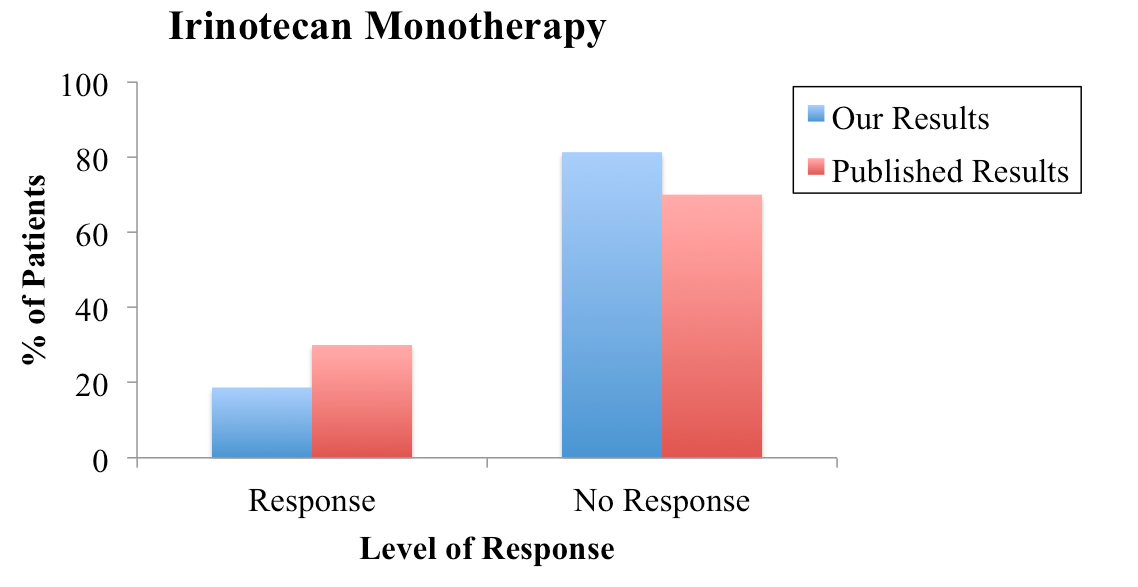}
\includegraphics[height=.2\textheight]{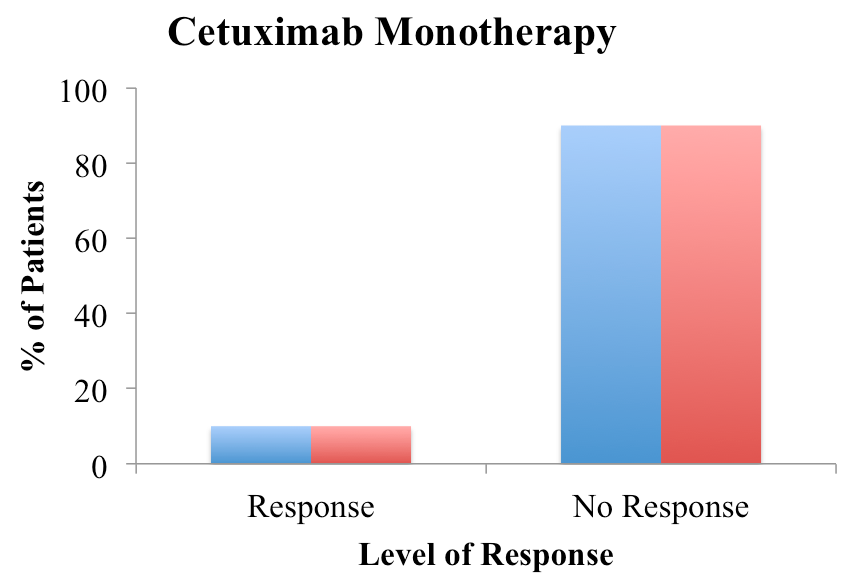}
\includegraphics[height=.2\textheight]{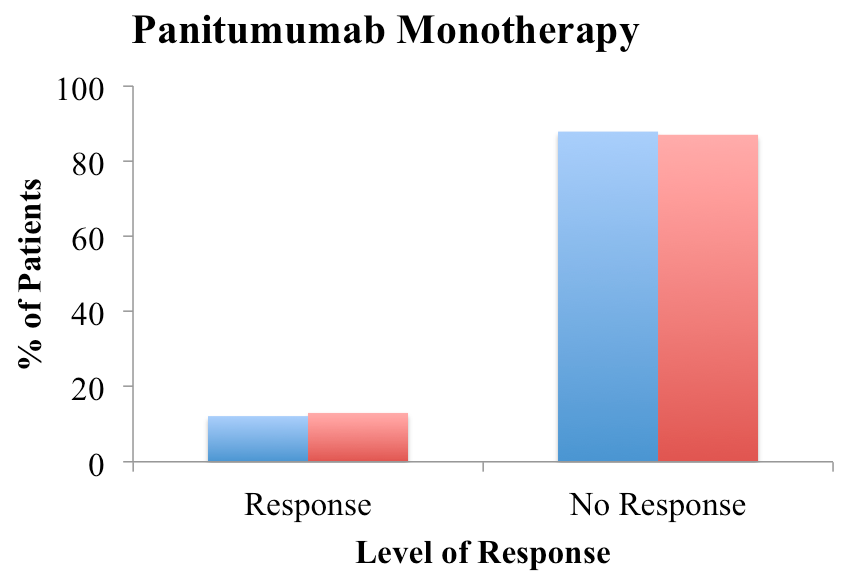}
\end{center}
\caption{{\bf Our clinical trial simulations compared to reported clinical trial results for irinotecan monotherapy (A), cetuximab monotherapy (B), and panitumumab monotherapy(C).} Our simulation results closely match published results for both cetuximab and panitumumab montherapies. For irinotecan monotherapies, the reduced response seen in our simulations is intended, since the patients receiving mAb therapy are often not as responsive as most patients to other treatments.}
\label{fig:monotherapies}
\end{figure}

	Monotherapy clinical trial simulations were performed for each of the three drugs used in our model. An irinotecan monotherapy clinical trial was simulated, using a common treatment regimen, and results were compared with clinical data.  (The treatment details can be  found in the ``Treatments'' section of the Supplementary Materials).
 Irinotecan monotherapy simulations resulted in a total response of 18.7\%, (15.6\% PR, 3.1\% CR), versus an overall reported response rate of 30\% (see Figure~\ref{fig:monotherapies}(A)). 
This is consistent with practice, since patients getting mAb treatments are often not very responsive to chemotherapy.
Cetuximab and panitumumab clinical trials were also simulated, using the common treatment regimens found in ``Treatments'', to verify that the desired response rate was achieved. Parameter calculations for each mAb drug involved choosing a value for $\psi$ that resulted in accurate clinical trial results for the mAb monotherapies, but verification of these values is important. Cetuximab monotherapy simulations matched the expected results with a total response rate of 10\%, (10\% PR, 0\% CR), versus an overall reported response rate of approximately 10\% (see Figure~\ref{fig:monotherapies}(B)), and panitumumab monotherapy simulations matched the expected results with a total simulation response rate of 12.15\% (10.9\% PR, 1.25\% CR), versus an overall reported response rate of 10-13\%  (see Figure~\ref{fig:monotherapies}(C)). Combination therapies, using either irinotecan and cetuximab or irinotecan and panitumumab, were also simulated. These simulations used the common treatments for each drug and gave the two treatments simultaneously. 

We do not currently have a way to adjust severity classification for the tumor based on patient health.
A smaller tumor in a very sick patient can be just as dangerous as a larger tumor in healthier patient.
Therefore, when examining monotherapies, which do not have particularly damaging effects on the immune system, we
measured responses after one week in order to capture the less dramatic and potentially transient effects, which could still
be helpful to patients whose immune systems have not been severely compromised by treatments.  
However, in the case
of combination treatments, we chose to wait longer after treatment before measuring results.  
The clinical trial studies summarized in Table \ref{tab:response_rates_common} did not report when tumor was measured after the last treatment, 
so we chose to measure tumor size four weeks after the final treatment for all simulations.
Although many more patients
experienced an initial drop in tumor size as a result of the combination treatments, this drop was frequently unhelpful to the patient because of the
additional loss of immune strength associated with the harsher combination treatments.

\begin{figure}[!ht]
\begin{center}
\includegraphics[height=.2\textheight]{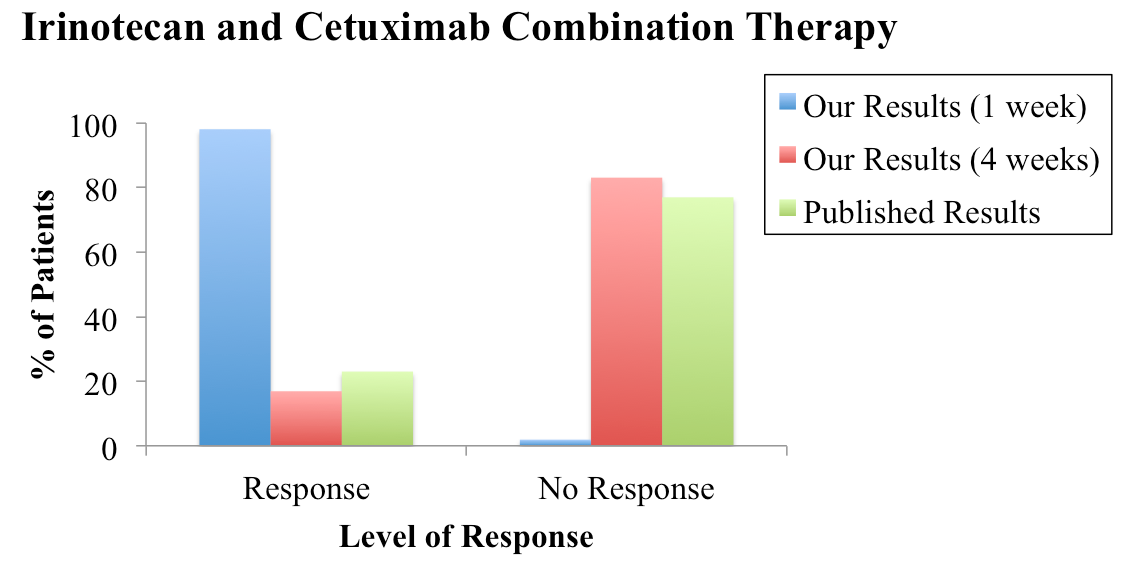}
\includegraphics[height=.2\textheight]{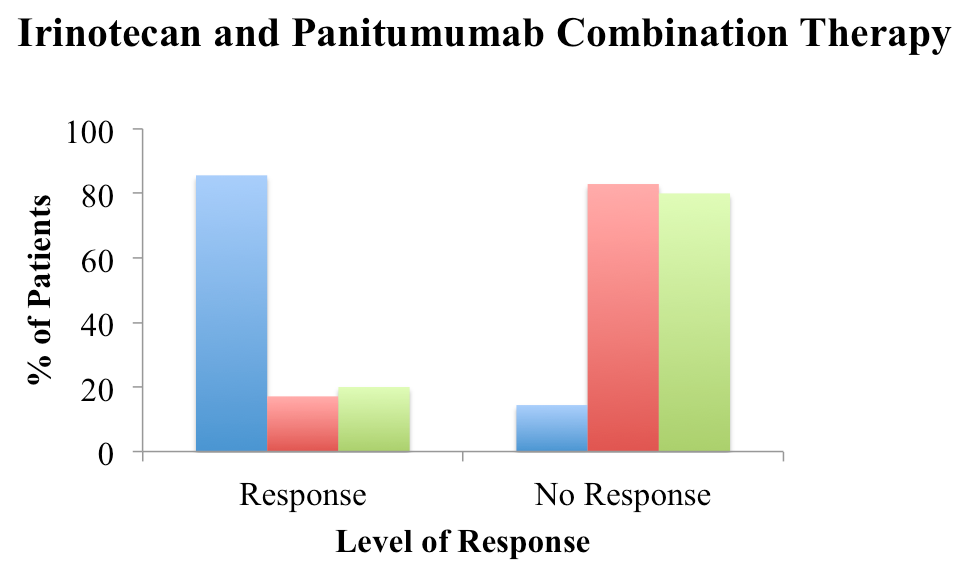}
\end{center}
\caption{{\bf Our clinical trial simulations compared to reported clinical trial results for irinotecan and cetuximab combination therapy (A) and irinotecan and panitumumab combination therapy (B).} If simulation results are measured one week post-treatment, they do not replicate published results for combination therapies. However, if simulation results are measured four weeks post-treatment, our results are very similar to published results.}
\label{fig:tx_combos}
\end{figure}

Our simulations match reported clinical trial results fairly closely (see Figure~\ref{fig:tx_combos}). The results from these simulations are also provided 
in Table~\ref{tab:response_rates_common},
along with the associated clinical trials data.

\subsection*{Impact of Patient Specific Response Parameters on Treatments}

\begin{figure}[!ht]
\begin{center}
\includegraphics[height=.2\textheight]{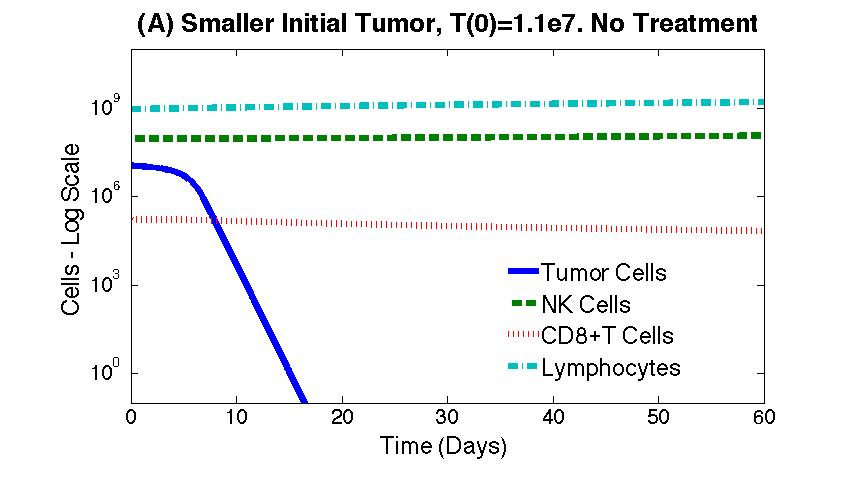}
\includegraphics[height=.2\textheight]{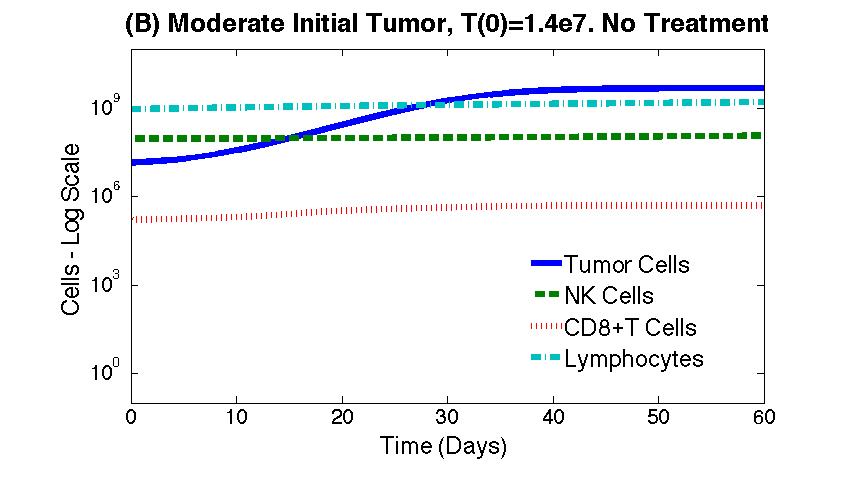}
\end{center}
\caption{{\bf Patients can end up at either the \emph{no tumor equilibrium} or the \emph{large tumor equilibrium}.} In (A) $T(0)=1.1\times10^7$ cells, in (B) $T(0)=1.4\times10^7$ cells.
A tumor with a small initial size will quickly shrink toward zero, a tumor with a larger initial size 
will quickly grow to the carrying capacity of the system. 
All other initial values are the same in both simulations.
}
\label{fig:stability}
\end{figure}

	We also ran the model 
to simulate individual patients, using set values for the patient-specific parameters, to examine how the tumor and immune system interact with strong or weak responses to the medications. The results from these simulations were plotted as cell populations/concentrations versus time. 
In Figure \ref{fig:stability}, we first 
see how the initial tumor size determines whether the tumor ultimately shrinks or
grows to carrying capacity in the absence of treatment.  In our remaining simulations that
include treatment,
we ensure that the initial tumor size is chosen to be sufficiently large so that it would
grow to carrying capacity in the absence of treatment.

\begin{figure}[!ht]
\begin{center}
\includegraphics[height=.28\textwidth]{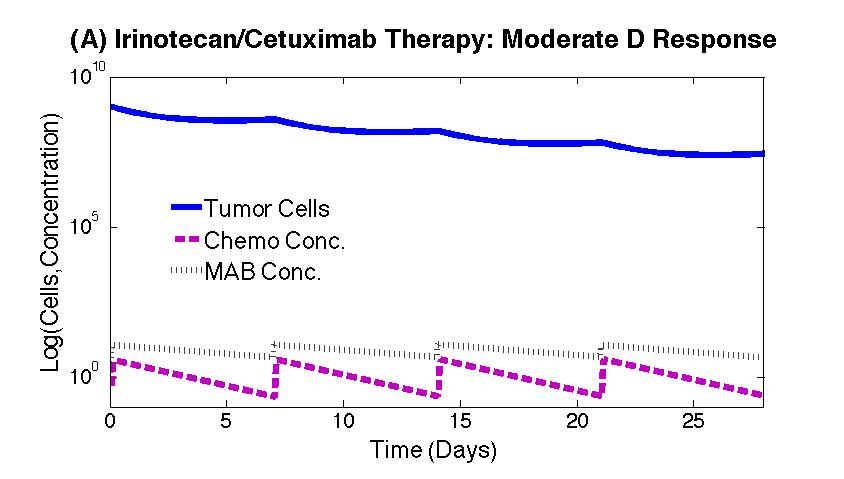} 
\includegraphics[height=.28\textwidth]{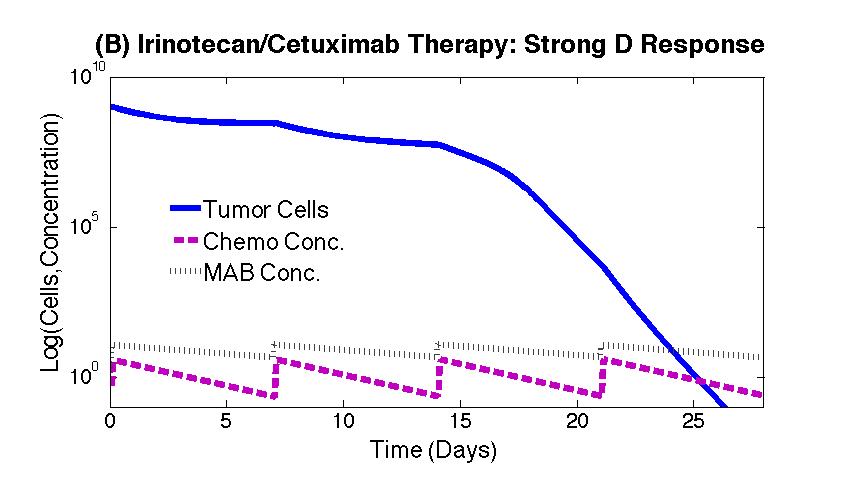} 
\end{center}
\caption{{\bf Effect of patient-specific immune response function $D$ on tumor 
response to irinotecan/cetuximab treatment.}  
Tumor response to irinotecan/cetuximab combination therapy with $l=1.6$ and $s=7\times10^{-3}$ (A), 
resulting in a moderate response $D$, and with $l=1.3$ and $s=4\times10^{-3}$ (B), resulting in a stronger 
response $D$. With the moderate $D$, the tumor will increase to carrying capacity with the 
cessation of treatment, but the stronger $D$ allows the patient's immune response to eradicate the tumor.}
\label{Deffect}
\end{figure}

In Figure \ref{Deffect}, 
we simulate irinotecan/cetuximab combination therapy, and can see how a
modification in an individual's CD8+T cell response to tumor, via response function $D,$ 
affects treatment outcomes.  
We also simulated the tumor response to irinotecan/panitumumab combination therapy (figure
not shown)
with $l=1.6$ and $s=7\times10^{-3},$ resulting in a moderate response $D$, 
and with $l=1.3$ and $s=4\times10^{-3},$ resulting in a high response $D.$ 
With the moderate $D$, the tumor will increase to carrying capacity 
with the cessation of treatment, but the stronger $D$ allows the patient's immune response to
eradicate the tumor.

\begin{figure}[!ht]
\begin{center}
\includegraphics[width=.48\textwidth]{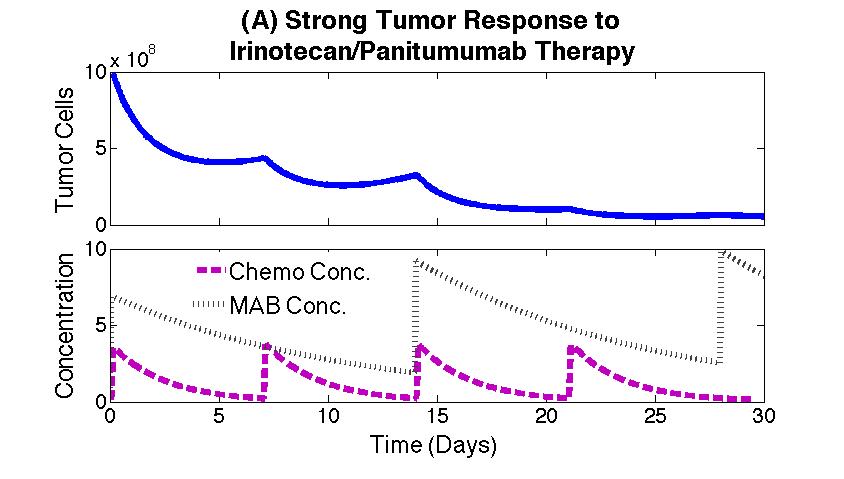}
\includegraphics[width=.48\textwidth]{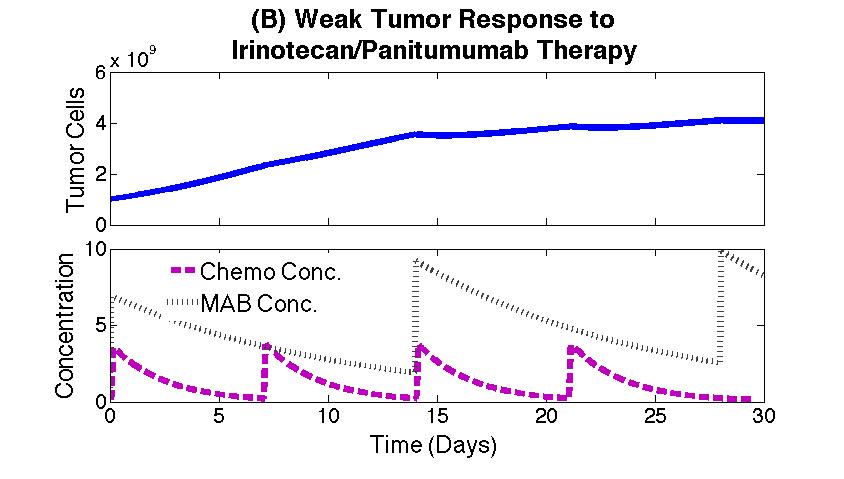}
\includegraphics[width=.48\textwidth]{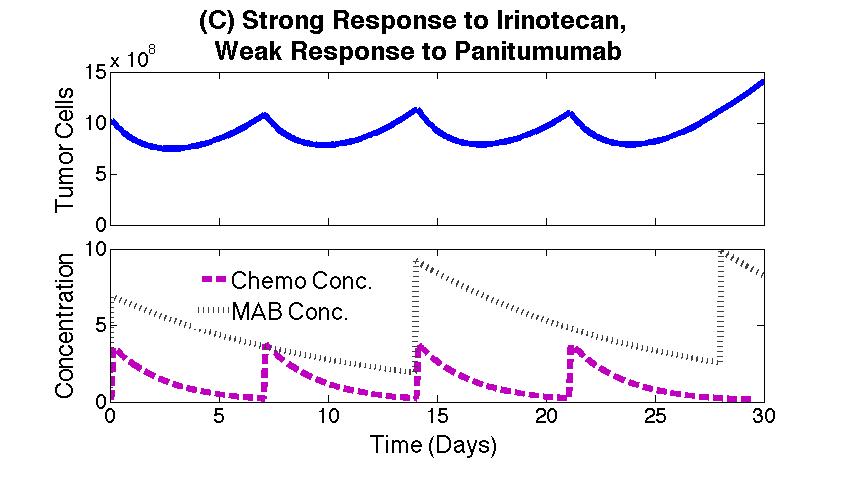}
\includegraphics[width=.48\textwidth]{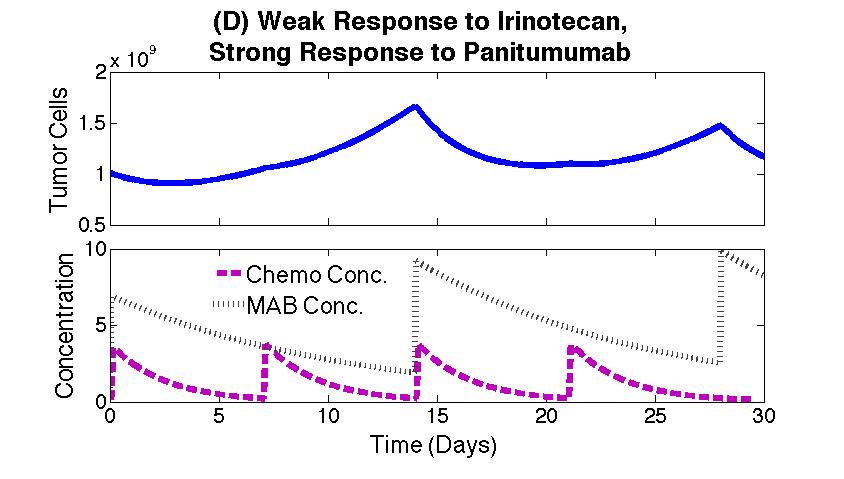}
\end{center}
\caption{{\bf Tumor responses to combination therapy with irinotecan and panitumumab.} When the tumor has a strong response (high $K_T$ and $\psi$, 100\% strength) to both medications (A), the tumor shrinks during the course of the treatment. When the tumor has a weak response (low $K_T$ and $\psi$, 10\% strength) to both medications (B), the tumor grows toward the carrying capacity. When the tumor has either a strong response to irinotecan and a weak response to panitumumab (C) or a weak response to irinotecan and a strong response to panitumumab (D), the tumor will fluxuate in size, but will stay approximately the same size overall during the treatment course.}
\label{TumorTreatmentResponse}
\end{figure}

In Figure \ref{TumorTreatmentResponse}, 
we observe how individual tumor sensitivity to
either chemotherapy or mAb therapy affects tumor size.  In particular, we simulate
four possible combinations: a strong response
to both chemo and mAb therapy (A), a weak response to both chemo and mAb (B), a strong response to chemo
but a weak response to mAb (C), and a strong response to mAb but a weak response to chemo (D).


\begin{figure}[!ht]
\begin{tikzpicture}[node distance = 3in, auto]
\node [label = {above: A: Weak Immune Response}] (A){ \includegraphics[width = .5\textwidth]{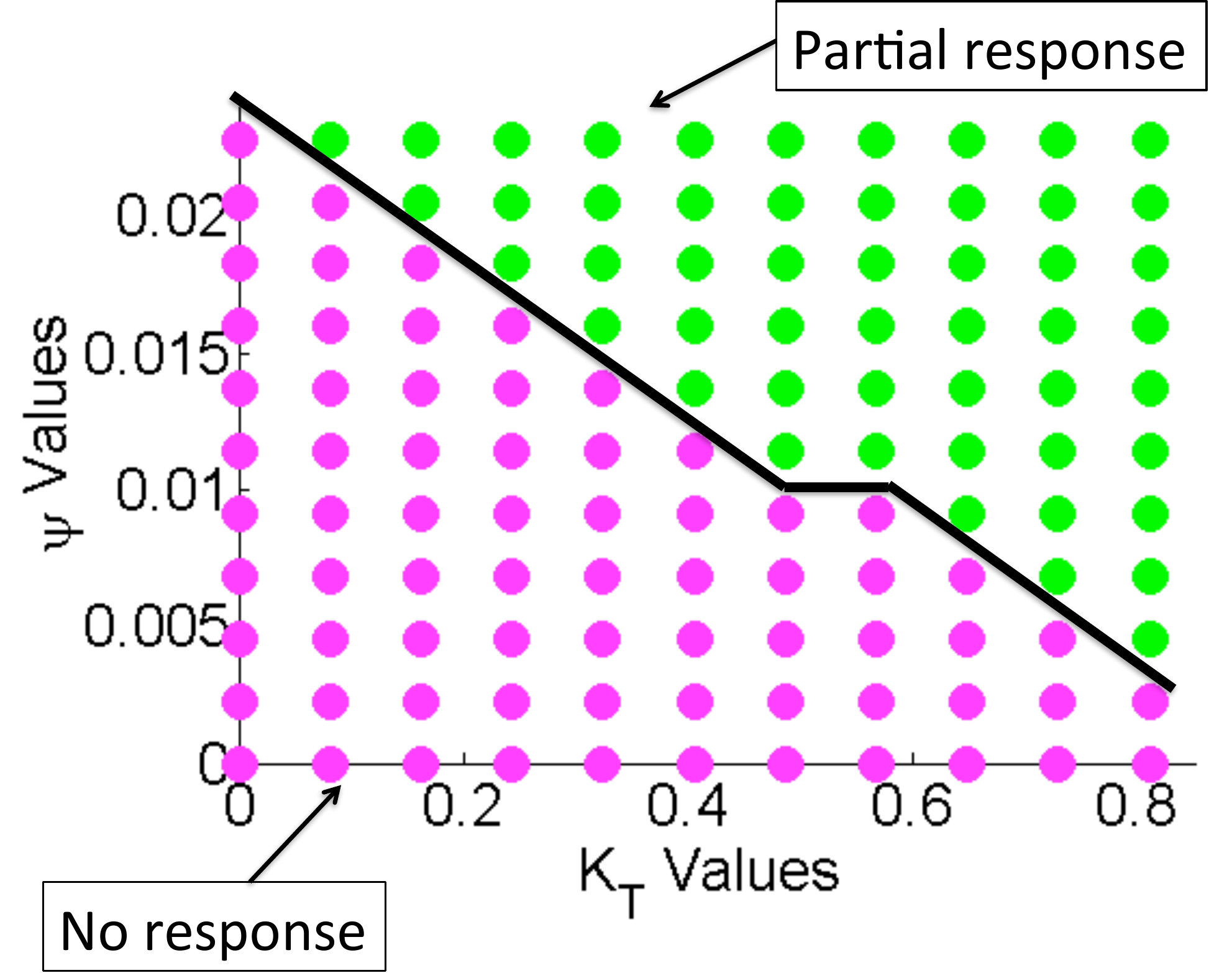}};
\node [right of = A, label = {above:B: Moderate Immune Response}](B){\includegraphics[width = .5\textwidth]{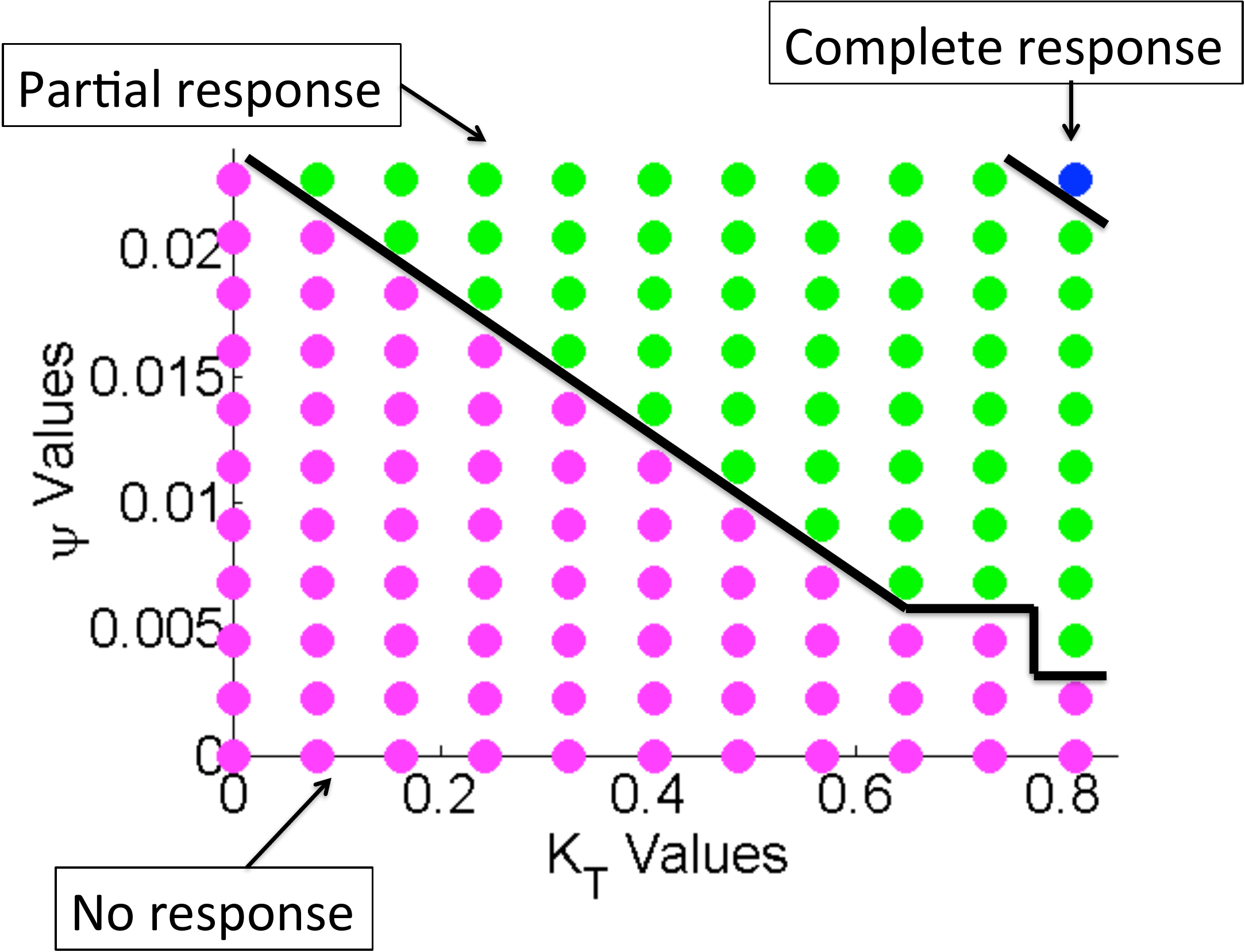}};
\node [label = {above:C: Strong Immune Response}](C) at (3,-7) {\includegraphics[width = .5\textwidth]{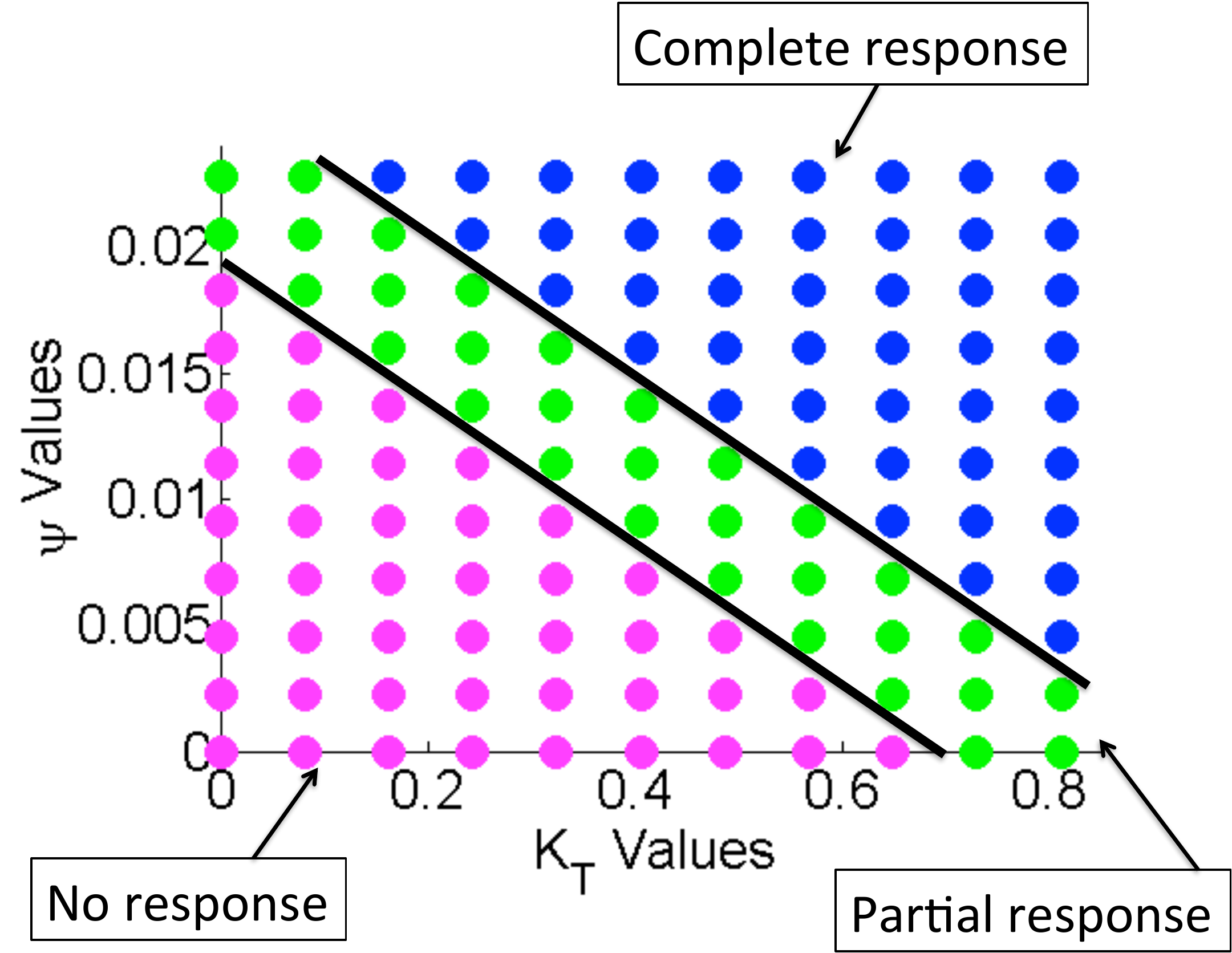}};
\end{tikzpicture}
\caption{{\bf Sensitivity to the strength of $\psi$ ($\in0-2.28$ L mg$^{-1}$Day$^{-1}$, examined for cetuximab only) and $K_T$ ($\in0-0.81$ Day$^{-1}$), for a patient with a weak (A), moderate (B), and strong (C) immune response .} \textcolor{blue}{Blue} means complete response to medication, \textcolor{green}{green} means partial response to medication, \textcolor{magenta}{magenta} means no response to medication.  In a patient with a weakened immune system, the medications will not be able to completely remove the tumor, even when the tumor cells have a maximal response to both medications. However, a patient with a strong immune system has a good chance of eliminating the tumor, as long as the patient's tumor cells have some response to the medications.
Parameter values used for the immune kill rate, $D$ are as follows. Weak response: $ d=1.3$, $l=2$, $s=4\times10^{-2}$;
moderate response: $d=1.6$, $l=1.4$, $s=8\times10^{-3}$; strong response: $d=2.1$, $l=1.1$, $s=5\times10^{-3}$ .}
\label{fig:KtPsi_sens}
\end{figure}

In order to explore which patient-specific parameters play a role in whether a patient will respond
to treatment,
	the effect of the variable parameters, $d$, $l$, $s$, $K_T$, and $\psi$, was also examined. $d$, $l$, 
and $s$ were fixed at three sets of values from the set of patient-specific parameters used for clinical trial 
modeling, a ``weak D" ($ d=1.3$, $l=2$, $s=4\times10^{-2}$);
``moderate D'' ($d=1.6$, $l=1.4$, $s=8\times10^{-3}$);and ``strong D'' ($d=2.1$, $l=1.1$, $s=5\times10^{-3}$ )
response. The variables $K_T$ and $\psi$ were then 
varied over their range of $0$ to their maximum values, using cetuximab as the mAb drug, and the model 
was run for 28 days with each pair of values. Figure~\ref{fig:KtPsi_sens}(A) shows that a patient with 
a weak inherent immune system cannot have a complete response, even with a full-strength response by 
the tumor to the chemotherapy and mAb treatments. A strong response by the tumor to either treatment will 
result in a partial response for the tumor overall. Figure~\ref{fig:KtPsi_sens}(B) shows that a 
patient with a moderately strong immune system has a chance of overpowering the tumor and obtaining a 
complete result, with strong tumor responses by the tumor to both the chemotherapy and mAb drugs. 
The patient is more likely however to have a partial overall response, resulting from a strong response 
by the tumor to only one medication, or to have no response. Figure~\ref{fig:KtPsi_sens}(C) shows that a 
patient with a strong immune system has a good chance for a complete overall response, with a strong 
response by the tumor to either the mAb or chemotherapy treatments. However, the patient will still not 
respond to the treatment if the tumor is only weakly affected by both of the medications.

\subsection*{Clinical Trial Simulations for Hypothetical Treatments}
\begin{figure}[!ht]
\begin{center}
\includegraphics[height=.60\textheight]{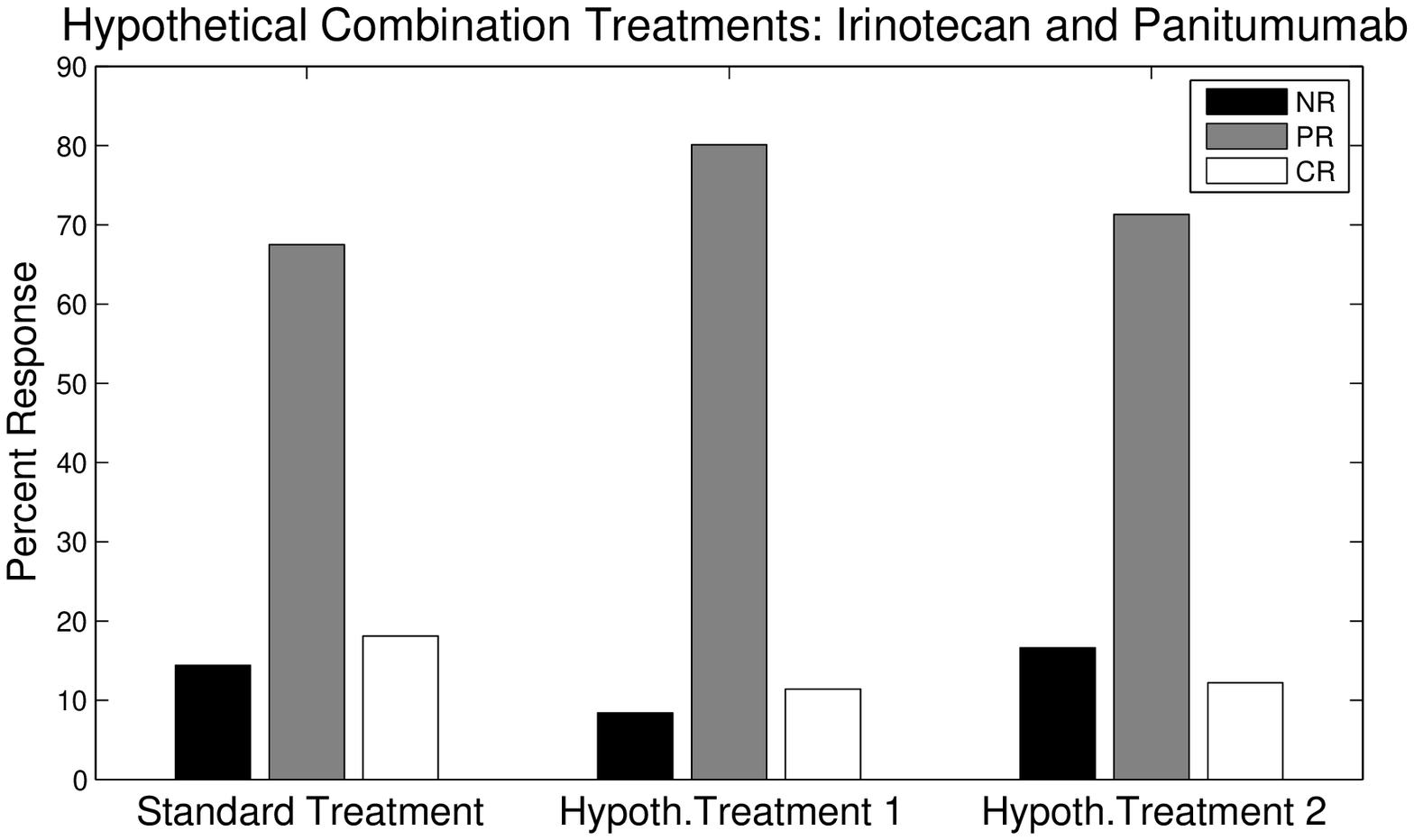} 
\includegraphics[height=.25\textheight]{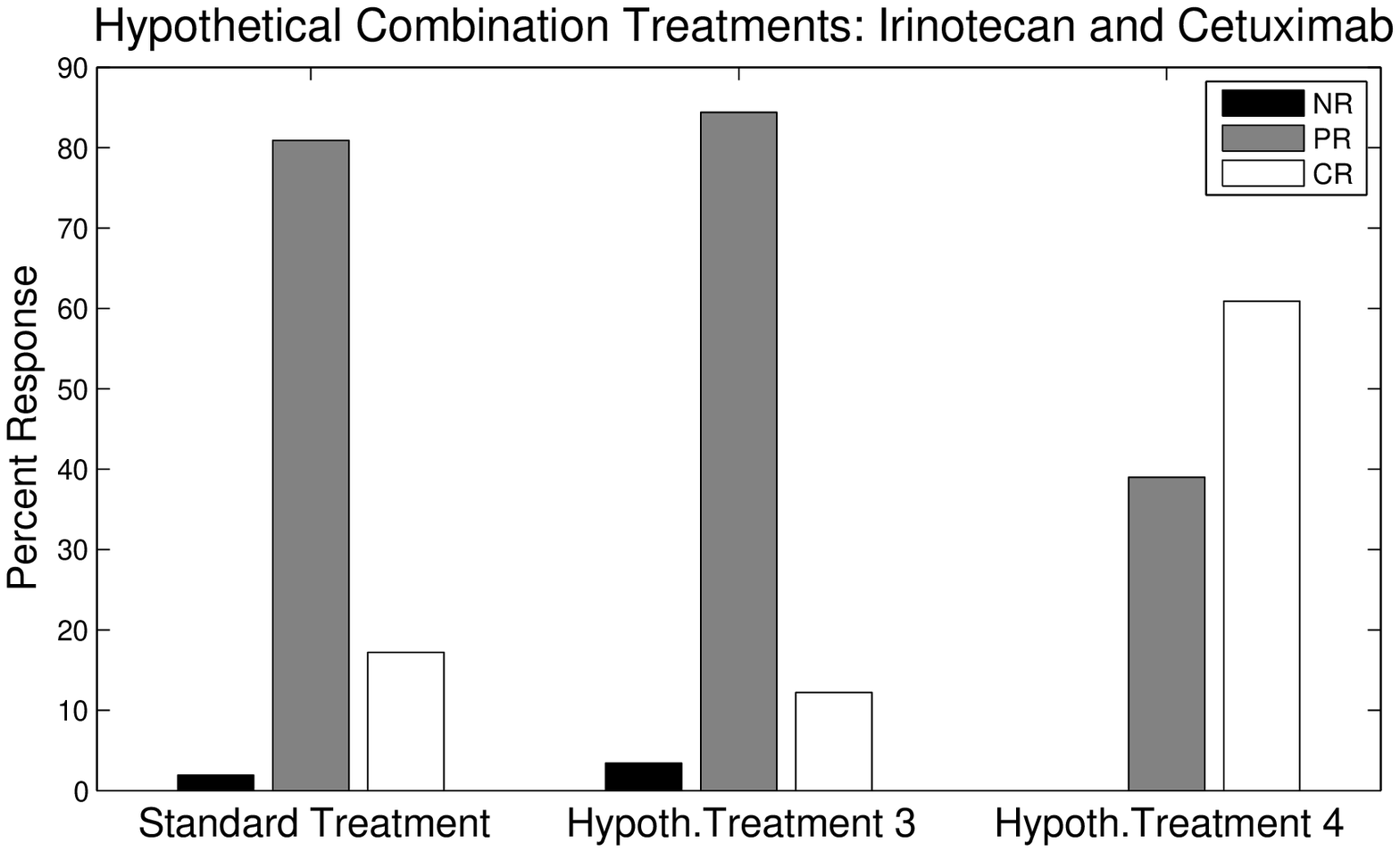}
\end{center}
\caption{Response rates from clinical trial simulations, comparing standard treatment to two experimental treatment schedules.  
Top panel is irinotecan and panitumumab.  
Bottom panel is irinotecan and cetuximab.  
NR, No Response. PR, Partial Response. CR, Complete Response.
320 individuals simulated. 
Dosing details in Table \ref{tab:response_rates_my_tx}.
}
\label{hypoth_combos}
\end{figure}

	Clinical trial simulations with hypothetical treatment combination regimens were also performed.  We explored various timings and dosing levels
of irinotecan in combination with cetuximab, and separately, irinotecan in combination with panitumumab. 
 Many of the combination treatments we experimented with, which used different doses, dosing frequencies, and different start times for each medication, were not as successful at shrinking the tumor as the current standard treatments. However, we did find 
some treatment regimens which appear to result in a smaller final tumor size, one with each of the mAb medications. 
These results are shown in Figure~\ref{hypoth_combos}.
For comparison, 
we include one set of
simulation results for tested treatments that can also be found in Table \ref{tab:response_rates_common},  
as well as the
results of the two separate hypothetical dosing schedules.
In Figure \ref{hypoth_combos}, top panel, we compare population responses to three different combination doses of irinotecan combined with panitumumab,
and in the bottom panel, we compare irinotecan combined with cetuximab.

Hypothetical Treatment 1:
One hypothetical treatment improvement can be seen when using irinotecan combined with panitumumab, required no change in dosing levels, 
but a change in the timing of the dose administration. In this case, we dose first with panitumumab, and then wait four days 
to begin irinotecan doses. Irinotecan is then continued every 7 days for the remainder of the treatment, while panitumumab continues to be administered
once every two weeks, as with a standard dosing schedule. This treatment decreased the total number of patients who did not respond to treatments 
from 14.4\% to 8.4\%, although it also decreased the number of patients who demonstrate a 
complete response from 18.1\% to 11.4\%. 
Simulation results can be seen in 
Figure~\ref{hypoth_combos}, top panel.
Since the medications are not being given at the same time, the patient may experience fewer simultaneous side 
effects with this treatment schedule. However, the treatment also requires the patient to make extra trips 
to the hospital for treatment administration.

Hypothetical Treatment 2:
In Figure~\ref{hypoth_combos}, top panel, we also show a second hypothetical treatment, 
in which the doses of both irinotecan and panitumumab are increased:
The irinotecan dose is 2.8 times the standard dose, and panitumumab is 1.5 times the standard dose.  However, dosing frequency is decreased to once
every three weeks for both medications.  This results in a slightly higher complete response rate of 12.2\%, and a partial response rate of 71.3\%.

Hypothetical Treatment 3:
In the third hypothetical scenario, shown in Figure~\ref{hypoth_combos}, bottom panel, we look at irinotecan combined with
cetuximab.   In this case, we modify the dose timing only, and leave dose
amounts at standard levels.  We
dose first with irinotecan, and follow up with a
cetuximab dose four days later.
This strategy was not particularly successful. The complete response rate
was only 12.2\%, as opposed to the 17.2\% achieved by the standard dosing
schedule.

Hypothetical Treatment 4:
Treatment option 4 combines a higher dose of irinotecan and a higher dose of 
cetuximab, both administered less frequently than standard treatment would
require.  Results are pictured in Figure~\ref{hypoth_combos}, bottom panel. Irinotecan is administered 
once every three weeks, and cetuximab is administered once every two weeks.
Treatment lasts nine weeks, so the individual receives three irinotecan doses,
and four cetuximab doses.
The use of these drugs at the higher doses, at least as monotherapies, has
been reported in the literature
\cite{gravalos_et_al_2009,sobrero_et_al_2008,lenz_2007}. 
The higher dosed irinotecan/cetuximab combination increases the overall
response rate from 98.9\% for the standard treatment to 100\%, and increases
the complete response rate from 17.2\% to 60.9\%.

Of all four hypothetical treatments presented, the high-dose
irinotecan/cetuximab combination appears to be the most effective. 
In our simulations, the lymphocyte count stayed above a specified minimum, 
which is one way to measure the degree of immune system damage from the
chemotherapy. 
With this treatment schedule, the
medications are not always given in the same weeks, which has the benefit of
the tumor population being kept low with frequent medications, while side
effects for the patient may be reduced. However, this treatment schedule also
requires that the patient receive medication every week, which may be an
inconvenience (versus, for example, the treatment with irinotecan and
panitumumab being given only every 3 weeks).

\subsection*{Sensitivity to Parameters}
	Parameter sensitivity analysis was performed to determine which model parameters have the greatest effect on tumor size, both in the absence of treatment and with different treatments.  We found seven parameters that significantly
affected tumor size in our simulations. In order to separate short term and long term effects, we looked
at tumor size seven days after initiation of the simulation, and again at twenty eight days after initiation. In
most cases, parameters that had a significant impact on tumor size at day seven were also significant at day twenty eight.
A full description of the parameters and their values can be found in the supplementary information, 
and in Tables~\ref{tab:tumor}-\ref{tab:meds}, but we will briefly explain here the parameters found to be most 
significant.

\begin{figure}[!ht]
\begin{center}
\includegraphics[height=.18\textheight]{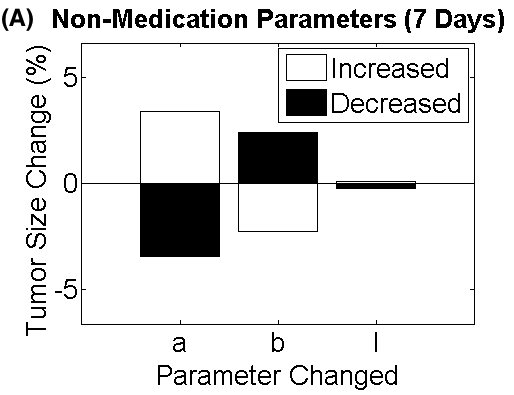}
\includegraphics[height=.18\textheight]{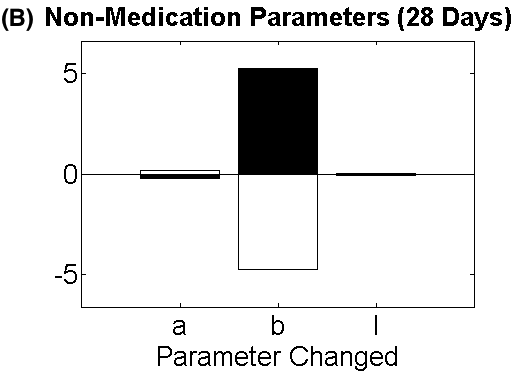}\\
\includegraphics[height=.18\textheight]{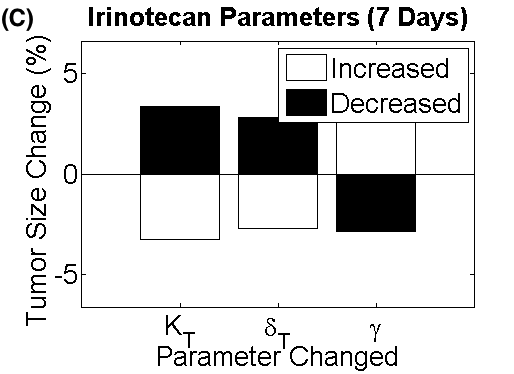}
\includegraphics[height=.18\textheight]{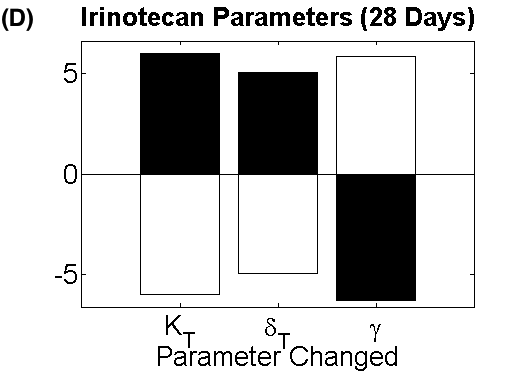}\\
\includegraphics[height=.18\textheight]{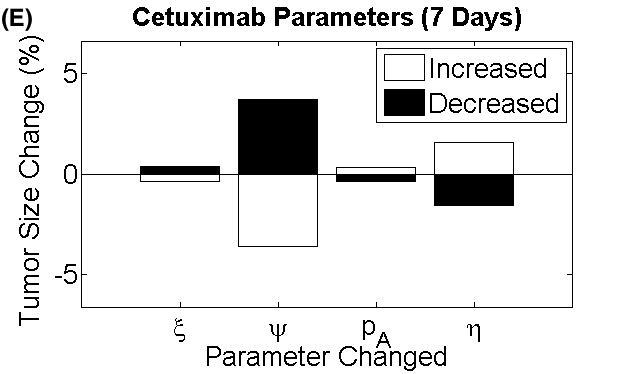}
\includegraphics[height=.18\textheight]{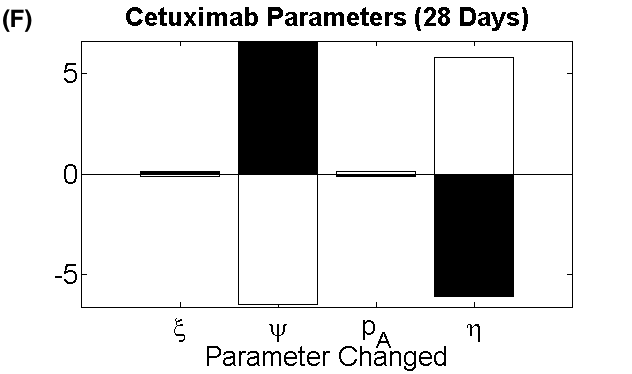}\\
\includegraphics[height=.18\textheight]{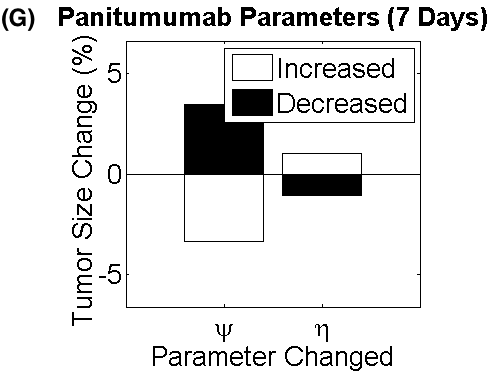}
\includegraphics[height=.18\textheight]{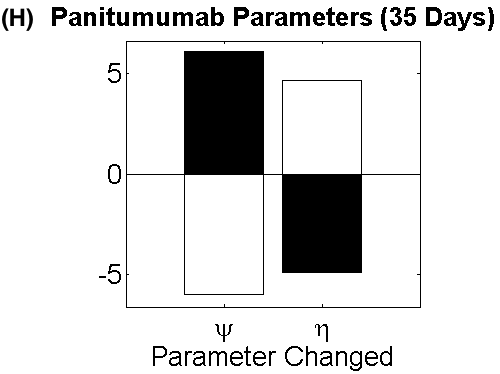}
\end{center}
\caption{{\bf Sensitivity of final tumor size to a 5\% change in parameters.} Final tumor size was measured at 7 days to capture short-term sensitivity and at 28 days (no medication, irinotecan parameters, and cetuximab parameters) or 35 days (panitumumab parameters) to capture sensitivity after treatments are completed. With no medication, final tumor size is most sensitive to exponential growth rate ($a$) and carrying capacity ($b$), at 7 days (A) and carrying capacity ($b$) at 28 days (B). Of the parameters for irinotecan treatments, final tumor size is most sensitive to irinotecan-induced tumor cell death rate ($K_T$), efficacy ($delta_T$), and the elimination rate ($\gamma$) at both 7 days (C) and 28 days (D). Of the parameters for cetuximab treatments, final tumor size is most sensitive to the rate of cetuximab-induced tumor death ($\psi$) and the elimination rate ($\eta$) at both 7 days (E) and 28 days (F). Of the parameters for panitumumab treatments, final tumor size is most sensitive to the rate panitumumab-induced tumor death ($\psi$) at 7 days (G) and both the rate panitumumab-induced tumor death ($\psi$) and the elimination rate ($\eta$) at 35 days (H). Parameters resulting in $<$0.05\% change in final tumor size with a 5\% change are not shown.}
\label{fig:param_sens}
\end{figure}

Each parameter value was individually increased and decreased by 5\% while all other parameter values were held 
constant. Tumor size was measured at 7 days, when the tumor is still growing very quickly in our model, and at 28 days, when it is close to its maximum volume in our model. 
First, 
we analyzed parameter sensitivity
in simulations with no treatments given, so treatment-related parameters did not affect simulation outcomes.
Results for parameters with the most significant impact on outcomes are shown in 
Figure~\ref{fig:param_sens}(A) and (B) at days 7 and 28. 
Note that, while $b$ (which represents the inverse of the carrying capacity) is by far the most important parameter in determining final tumor size, $a$ (the intrinsic tumor growth rate) is important in determining how quickly the 
tumor reaches its maximum volume. 
The parameter $l$, 
which affects the functional form of the CTL kill rate,
has the most significant effect on non-medicated initial tumor growth of all the immune system parameters.

	A sensitivity analysis with treatment-related parameters was then performed.
For chemotherapy irinotecan treatment parameters, the final tumor size was found to be very sensitive to $K_T$ and $\delta_T$, which determine the model's response to the chemotherapy drug, and to $\gamma$, which represents the excretion of the chemotherapy drug (see Figure~\ref{fig:param_sens}(C) and (D)). Tumor regrowth between treatments was much more dependent on $\gamma$ than was the decrease in tumor size following treatments. This makes sense, 
because when the chemotherapy remains in the body longer, it will be more effective at maintaining lower tumor volumes between treatments.

We next tested the monoclonal antibody therapies, cetuximab and panitumumab, separately. 
Dose timings for cetuximab and panitumumab are different, so we measured
parameter sensitivity 
according to the different lengths of a standard course of treatment for each treatment type.
For cetuximab,  we consider one course of treatment to be on days 0, 7, 14 and 21 (four treatments
total, once per week over four weeks), whereas for
panitumumab, one course of treatment is on days 0, 14, and 28 (three treatments total,
once every other week for three weeks).  We then measured tumor size one week after the
last dose of the treatment course.  Therefore, long term sensitivity for cetuximab treated tumors 
was measured at day 28, and for panitumumab at day 35.
In both cases, the final tumor size was 
found to be sensitive to $\psi$, the strength of the tumor's response to mAb drugs, and to $\eta$, the mAb turnover rate (see Figure~\ref{fig:param_sens}(E-H)). 
This is reasonable, since the main anti-tumor activity of mAb medications is through interference with the ability of EGF to bind to EGFR on the tumor cell surface, an activity which is included in the term $\psi$~\cite{cancer_tx_book}. In the short term, cetuximab also shows some sensitivity to $\xi$ and $p_A$, which are the parameters that determine the strength of ADCC activity.


We note that a five percent change in all the remaining paramaters negligibly affected final tumor size.
In particular, the parameter $K_{TA}$,  which represents the 
increase in effectiveness of chemotherapy when it is used in conjunction with mAb therapy, 
had very little effect on final tumor size.
In this case, the final tumor size after 
28 days changed by less than 0.5 percent with a five percent change in $K_{TA}$ (figure not shown).


\section{Discussion}

We have extended the mathematical model presented in ~\cite{de_pillis_et_al_2009}
to include monoclonal antibody treatment. 
We have tuned the parameter values of the model to make them specific to
colorectal cancer, the chemotherapy treatment irinotecan, and the monoclonal antibody treatments
cetuximab and panitumumab. 
Two stable equilibrium states were found numerically, 
a \emph{no tumor equilibrium} and a \emph{large tumor equilibrium}. 
Tumors can be driven to either of these states in simulations, depending on the relative 
strength of the patient's immune system and the treatments given. 

Colorectal tumors can have a wide variety of mutations, and some of these mutations 
limit a medication's ability to function fully.
The parameters $K_T$ and $\psi$  represent a range 
of different tumor responses to the same chemotherapy and mAb treatments. 
At the beginning of a simulation for an individual, values for these parameters can
be chosen randomly from within proscribed biological ranges.
Use of these randomly chosen variables allows us to replicate the
population level results seen in 
clinical trials. 

A clinical study can be simulated by numerically solving the model 
multiple times to represent each individual outcome in the study.
In our simulations, we solved the model with 64 different combinations of patient parameters.
When simulating individuals receiving mAb 
monotherapy, the resulting population level response rates are quantitatively very close to the 
reported rates from clinical trials.

The simulation response rates for irinotecan chemotherapy was lower than the 
response rates reported in \cite{cancer_tx_book}. 
We intentionally chose model parameters to yield this outcome.
This is because we are assuming that our cohort of 64 individuals
have already have had chemotherapy with less success than would be seen
in a general population, and are
therefore in need of additional mAb therapy \cite{cunningham_et_al_2004}.  
On the other hand, 
the cohort in \cite{cancer_tx_book} was from the general population.


For combination therapies, our tumor population 
responded too well
to the medication short 
term, although in the long term, our simulated responses matched
experimental response rates well. 
The 
short-term over-responsiveness could be caused by inaccuracies in the model parameters 
or by time frame differences in the reported response rates. 
One possible inaccuracy in the model is that the variability in tumor responses to 
medication may not be accurately represented by the random variables. 
Tumors cells that aren't destroyed by one medication may be less likely to respond to 
another medication as well, such as cells in the center of the tumor, to which 
the medications would have limited access. Because mAbs and chemotherapy drugs 
generally have very different targets and mechanisms, a mutation causing the 
tumor to be refractory to one medication won't necessarily cause it to be 
refractory to the other, but it is possible. If that were the case, 
one model improvement might be to use 
one random variable to represent the tumor's response to both medications, 
instead of two variables. 
Most response rates are not actually reported with a time frame, so the response rates found 
with this model from four weeks post-treatment may be more consistent with real life measurements 
than the response rates from one week post-treatment. If this is the case, our model closely 
matches response rates for combination therapies as well, and the apparent 
over-responsiveness seen in our model with combination therapies is just caused by 
taking tumor measurements too early. Reported clinical trials for the dual 
treatments also often did not specify irinotecan dosing, whether the patients previously received 
treatment, or how long the treatment was given, so this may be responsible for part of the 
difference in response rates as well. Overall, our model gives a qualitatively good prediction 
of likely results for various dosing schedules.

In many of the experimental treatments, particularly in the high-dose treatments, the 
simulated individuals's immune system was also greatly weakened by the treatments, particularly
by irinotecan. 
Thus, although the tumor 
cell population was greatly reduced by the treatments, the individual's immune system 
was still unable to destroy the remaining tumor cells. 
Although we did not find 
much information about the
use of CD8+T-cell treatments for 
colorectal cancer in the literature, the addition of this treatment during the 
chemotherapy and mAb drug courses could help to bolster the immune system and allow 
the patient's immune response to 
lyse
tumor cells more effectively.  
This model, with the addition of a CD8+T cell treatment component, could be used to test 
this treatment hypothesis.

The parameter sensitivity analysis yielded results that were intuitively reasonable.  
The analysis also serves to highlight which parameters could be possible targets for reducing 
tumor size. For example, if we can get a better sense of biologically how to 
influence $l$, a parameter that affects the functional form of the CTL kill rate,
a large decrease in $l$ would result in an immune system that is able to conquer 
the tumor much more easily than the immune system resulting from a change in the other immune 
system parameters.

In the future, 
two modifications to this model could yield even more realistic outcomes.
First, we could tailor the parameters $K_T$ and $\psi$ to have a more specific biological 
meaning. For example, the KRAS mutation is known to be present in about 40\% of all colorectal tumors, 
and is known to reduce the effectiveness of mAb treatment to almost 
zero~\cite{de_roock_et_al_2008,amado_et_al_2008}. Information such as the EGFR counts on the 
tumor cells and the presence or absence of the KRAS mutation in an individual's 
tumor could allow for more personalized and specific parameter values, chosen 
from a smaller distribution based on features of the tumors cells, instead of from a 
larger random distribution. 
Second, an equation representing patient well-being could be very useful for predicting 
effective treatments. Although using lymphocyte count allows us to determine that 
the patient's immune system has not been completely destroyed by the medication, 
it doesn't take into account factors such as the inconvenience of frequent treatments,
the fact that high doses of cytotoxic medication
may result in 
side effects harmful to cells of the body other than immune cells (such as those of the stomach
lining).

There are several notable clinical observations that are important in informing the next stages of model development.
One of these is that tumor cells become resistant to chemotherapeutic drugs, making disease progression very sensitive to the timing and dosing used in treatments, \cite{Ballesta2014}.  The expansion of the model to include a tumor population resistant to a particular drug would allow {\it in silico} testing of a variety of treatment scenarios.  Another aspect of treatment to bear in mind is the effect of an individual's circadian fluctuations on the tumor's susceptibility  to cytotoxic agents.  These periodic fluctuations can be captured in our model by allowing time-varying parameters or by introducing delays into the model.  Some models of colon cancer that do include circadian rhythms are discussed in \cite{Ballesta2014} and in the references therein.

\clearpage 

\section{Tables}
\begin{center}
\begin{longtable}{llp{3in}c}
\caption{{\bf Tumor Equation Terms and Parameters.} }\\ \hline
&&&\\
\bf{Term}					&	\bf{Param}	&	\bf{Description and Value (Units)}					& \bf{Source}\\
\endfirsthead

\multicolumn{4}{c}%
{{\bfseries \tablename\ \thetable{} -- continued from previous page}} \\

\bf{Term}					&	\bf{Param}	&	\bf{Description and Value (Units)}					& \bf{Source}\\
\endhead

\multicolumn{4}{r}{{\em Table cont'd on next page.}}\\ \hline
\endfoot
\hline\hline
\endlastfoot

$aT(1-bT)$					&			&	Logistic tumor growth							& \\
							&	$a$		&	Growth rate of tumor							& \\
							&			&	$2.31\times10^{-1}$ (Day$^{-1}$)					& \cite{corbett_et_al_1975}\\
							&	$b$		&	Inverse of carrying capacity						& \\
							&			&	$2.146\times10^{-10}$ (Cells$^{-1}$)				& \cite{leith_et_al_1987}\\
$-cNT$						&			&	NK-induced tumor death							& \\
							&	$c$		&	Rate of NK-induced tumor death					& \\
							&			&	$5.156\times10^{-14}$ (L Cells$^{-1}$Day$^{-1}$)	& \cite{de_pillis_et_al_2009}\\
$-\xi\frac{A}{h_1+A}NT$			&			&	mAb-induced tumor death from NK cell interactions	& \\
							&	$\xi$	&	Rate of NK-induced tumor death through ADCC		& \\
							&			&	$6.5\times10^{-10}$ (L Cells$^{-1}$Day$^{-1}$)$^a$	& \cite{kurai_et_al_2007}\\
							&			&	$0$ (L Cells$^{-1}$Day$^{-1}$)$^b$				& \cite{grothey_2006}\\
							& 	$h_1$	& 	Concentration of mAbs for half-maximal increase in ADCC	& \\
							&			&	$1.25\times10^{-6}$ (mg L$^{-1}$)$^a$			& \cite{kurai_et_al_2007}\\
							&			&	$0$ (mg L$^{-1}$)$^b$							& \cite{grothey_2006}\\
$-DT$						&			&	CD$8^+$T cell-induced tumor death				& \\
							&	$d$		&	Immune-system strength coefficient				& \\
							&			&	$\{1.3,1.6,1.9,2.1\}$, (Day$^{-1}$)						& \cite{de_pillis_et_al_2009}\\
							&	$l$		&	Immune-system strength scaling coefficient			& \\
							&			&	$\{1.1,1.4,1.7,2.0\}$, (--)								& \cite{de_pillis_et_al_2009}\\
							&	$s$		&	Value of $(\frac LT)^l$ necessary for half-maximal CD$8^+$T-cell effectiveness against tumor	& \\
							&			&	$\{4\times10^{-3},7\times10^{-3},9\times10^{-3},3\times10^{-2}\}$, (L$^{-1}$) 							& \cite{de_pillis_et_al_2009}\\
$-K_T(1-e^{-\delta_TM})T$		&			&	Chemotherapy-induced tumor death				& \\
							&	$K_T$	&	Rate of chemotherapy-induced tumor death			& \\
							&			&	$0-8.1\times10^{-1}$ (Day$^{-1}$)				& \cite{vilar_et_al_2008}\\
							& $\delta_T$ 	&	Medicine efficacy coefficient						& \\
							&			&	$2\times10^{-1}$ (L mg$^{-1}$)					& \cite{vilar_et_al_2008}\\
$-K_{AT}A(1-e^{-\delta_TM})T$	&			&	mAb-induced tumor death from increase in chemotherapy effectiveness	& \\
							&	$K_{AT}$	&	Additional chemotherapy-induced tumor death due to mAbs	& \\
							&			&	$4\times10^{-4}$ (L mg$^{-1}$Day$^{-1}$)$^a$ 		& \emph{ad hoc} value\\
							&			&	$4\times10^{-4}$ (L mg$^{-1}$Day$^{-1}$)$^b$ 		& \emph{ad hoc} value\\
$-\psi AT$					&			&	mAb-induced tumor death						& \\
							&	$\psi$	&	Rate of mAb-induced tumor death					& \\
							&			&	$0-2.28\times10^{-2}$ (L mg$^{-1}$Day$^{-1}$)$^a$	& \cite{cancer_tx_book}\\
							&			&	$0-3.125\times10^{-2}$ (L mg$^{-1}$Day$^{-1}$)$^b$	& \cite{cancer_tx_book}\\
\label{tab:tumor}
\end{longtable}
\begin{flushleft}
Descriptions of the biological relevance of each term and parameter and the parameter values in $T(t)$, which tracks the tumor cell population.\\
$^a$ For cetuximab.\\
$^b$ For panitumumab.
\end{flushleft}
\end{center}

\begin{center}
\begin{longtable}{llp{2.5in}c}
\caption{\bf{NK Cell Equation Terms and Parameters.}} \\ \hline
&&& \\
\bf{Term}				&	\bf{Param}	&	\bf{Description and Value (Units)}						& \bf{Source}\\
\endfirsthead

\multicolumn{4}{c}%
{{\bfseries \tablename\ \thetable{} -- continued from previous page}} \\

\bf{Term}		&	\bf{Param}	&	\bf{Description and Value (Units)}			& \bf{Source}\\ 
\endhead

\multicolumn{4}{r}{{\em Table cont'd on next page.}}\\ \hline
\endfoot
\hline\hline
\endlastfoot

$eC$						&			&	Production of NK cells from circulating lymphocytes	& \\
							& $\frac ef$	&	Ratio of NK cell synthesis rate with turnover rate		& \\
							&			&	$\frac 19$ (--)									& \cite{de_pillis_et_al_2009}\\
$-fN$						&			&	NK turnover									& \\
							&	$f$		&	Rate of NK cell turnover							& \\
							&			&	$1\times10^{-2}$ (Day$^{-1}$)					& Modified from \cite{de_pillis_et_al_2009}\\
$-pNT$						&			&	NK death by exhaustion of tumor-killing resources	& \\
							&	$p$		&	Rate of NK cell death due to tumor interaction		&\\
							&			&	$5.156\times10^{-14}$ (L Cells$^{-1}$Day$^{-1}$)	& \cite{de_pillis_et_al_2009}\\
$-p_A\frac{A}{h_1+A}NT$		&			&	Additional NK death by exhaustion of tumor-killing resources from mAb interactions & \\
							&	$p_A$	&	Rate of NK cell death due to tumor-mAb complex interaction & \\
							&			&	$6.5\times10^{-10}$ (L Cells$^{-1}$Day$^{-1}$)$^a$	& \cite{kurai_et_al_2007}\\
							&			&	$0$ (L Cells$^{-1}$Day$^{-1}$)$^b$				& \cite{grothey_2006}\\
$\frac{p_NNI}{g_N+I}$			&			&	Stimulatory effect of IL-2 on NK cells				& \\
							&	$p_N$	&	Rate of IL-2 induced NK cell proliferation			&\\
							&			&	$5.13\times10^{-2}$ (Day$^{-1}$)					& \cite{de_pillis_et_al_2009}\\
							&	$g_N$	&	Concentration of IL-2 for half-maximal NK cell proliferation & \\
							&			&	$2.5036\times10^5$ (IU L$^{-1}$)					& \cite{de_pillis_et_al_2009}\\
$-K_N(1-e^{-\delta_NM})N$		&			&	Death of NK cells due to chemotherapy toxicity		& \\
							&	$K_N$	&	Rate of NK depletion from chemotherapy toxicity		& \\
							&			&	$9.048\times10^{-1}$ (Day$^{-1}$)				& \cite{catimel_et_al_1995}\\
							& $\delta_N$	&	Chemotherapy toxicity coefficient					& \\
							&			&	$2\times10^{-1}$ (L mg$^{-1}$)					& \cite{vilar_et_al_2008}\\
\label{tab:nk_cells}
\end{longtable}
\begin{flushleft}Descriptions of the biological relevance of each term and parameter and the parameter values in $N(t)$, which tracks the concentration of NK cells.\\
$^a$ For cetuximab.\\
$^b$ For panitumumab.
\end{flushleft}
\end{center}

\begin{center}
\begin{longtable}{llp{2.5in}c}
\caption{\bf{CD8+ T Cell Equation Terms and Parameters.}}\\ \hline
&&& \\

\bf{Term}		&	\bf{Param}	&	\bf{Description and Value (Units)}			& \bf{Source}\\
\endfirsthead

\multicolumn{4}{c}%
{{\bfseries \tablename\ \thetable{} -- continued from previous page}} \\
\bf{Term}		&	\bf{Param}	&	\bf{Description and Value (Units)}			& \bf{Source}\\
\endhead

\multicolumn{4}{r}{{\em Table cont'd on next page.}}\\ \hline
\endfoot
\hline\hline
\endlastfoot

$\frac{\theta mL}{\theta+I}$		&			&	CD$8^+$T-cell turnover								& \\
							&	$\theta$	&	Concentration of IL-2 to halve CD$8^+$T-cell turnover		&\\
							&			&	$2.5036\times10^{-3}$ (IU L$^{-1}$)					& \cite{de_pillis_et_al_2009}\\
							&	$m$		&	Rate of activated CD$8^+$T-cell turnover				& \\
							&			&	$5\times10^{-3}$ (Day$^{-1}$)						& Modified from \cite{de_pillis_et_al_2009}\\
$j\frac{T}{k+T}L$				&			&	CD$8^+$T-cell stimulation by CD$8^+$T cell-lysed tumor-cell debris & \\
							&	$j$		&	Rate of CD$8^+$T-cell lysed tumor cell debris activation of CD$8^+$ T cells & \\
							&			& $1.245\times10^{-4}$ (Day$^{-1}$)						& Modified from \cite{de_pillis_et_al_2009}\\
							&	$k$		&	Tumor size for half-maximal CD$8^+$T-lysed debris CD$8^+$T activation & \\
							&			& $2.019\times10^7$ (Cells)								& \cite{de_pillis_et_al_2009}\\
$-qLT$						&			&	CD$8^+$T-cell death by exhaustion of tumor-killing resources & \\
							&	$q$		&	Rate of CD$8^+$T-cell death due to tumor interaction		& \\
							&			&	$5.156\times10^{-17}$ (Cells$^{-1}$Day$^{-1}$)			& Modified from \cite{de_pillis_et_al_2009}\\
$r_1NT$						&			&	CD$8^+$T-cell stimulation by NK-lysed tumor-cell debris	& \\
							&	$r_1$	&	Rate of NK-lysed tumor cell debris activation of CD$8^+$T cells & \\
							&			&	$5.156\times10^{-12}$ (Cells$^{-1}$Day$^{-1}$)			& \cite{de_pillis_et_al_2009}\\
$r_2CT$						&			&	Activation of natvie CD$8^+$T cells in the general lymphocyte population & \\
							&	$r_2$	&	Rate of CD$8^+$T-cell production from circulating lymphocytes & \\
							&			&	$1\times10^{-15}$ (Cells$^{-1}$Day$^{-1}$)				& Modified from \cite{de_pillis_et_al_2009}\\
$-\frac{uL^2CI}{\kappa + I}$		&			&	Breakdown of surplus CD$8^+$T cells in the presence of IL-2 & \\
							&	$u$		&	CD$8^+$T-cell self-limitation feedback coefficient		& \\
							&			&	$3.1718\times10^{-14}$ (L$^2$Cells$^{-2}$Day$^{-1}$)	& \cite{de_pillis_et_al_2009}\\
							&	$\kappa$	&	Concentration of IL-2 to halve magnitude of CD$8^+$T-cell self-regulation & \\
							&			&	$2.5036\times10^3$ (IU L$^{-1}$)						& \cite{de_pillis_et_al_2009}\\
$-K_L(1-e^{-\delta_LM})L$		&			&	Death of CD$8^+$T cells due to chemotherapy toxicity		& \\
							&	$K_L$	&	Rate of CD$8^+$T-cell depletion from chemotoxicity 		& \\
							&			& $4.524\times10^{-1}$	 (Day$^{-1}$)						& \cite{catimel_et_al_1995}\\
							& $\delta_L$	&	Chemotherapy toxicity coefficient						& \\
							&			& $2\times10^{-1}$ (L mg$^{-1}$)							& \cite{vilar_et_al_2008}\\
$\frac{p_ILI}{g_I+I}$				&			&	Stimulatory effect of IL-2 on CD$8^+$T cells				& \\
							&	$p_I$	&	Rate of IL-2 induced CD$8^+$T-cell activation			& \\
							&			&	$2.4036$ (Day$^{-1}$)								& \cite{de_pillis_et_al_2009}\\
							&	$g_I$	&	Concentration of IL-2 for half-maximal CD$8^+$T-cell activation & \\
							&			&	$2.5036\times10^3$ (IU L$^{-1}$)						& \cite{de_pillis_et_al_2009}\\
\label{tab:t_cells}
\end{longtable}
\begin{flushleft}Descriptions of the biological relevance of each term and parameter and the parameter values in $L(t)$, which tracks the concentration of CD8+ T cells.\\
\end{flushleft}
\end{center}

\begin{center}
\begin{longtable}{llp{3in}c}
\caption{\bf{Lymphocyte Equation Terms and Parameters.}} \\ \hline
&&&\\
\bf{Term}		&	\bf{Parameter}		&	\bf{Description and Value (Units)}			& \bf{Source}\\

\endfirsthead
\multicolumn{4}{c}
{{\bfseries \tablename\ \thetable{} -- continued from previous page}} \\

\bf{Term}		&	\bf{Parameter}		&	\bf{Description and Value (Units)}			& \bf{Source}\\
\endhead

\multicolumn{4}{r}{{\em Table cont'd on next page.}}\\ \hline
\endfoot
\hline\hline
\endlastfoot

$\alpha$						&			&	Lymphocyte synthesis in bone marrow					& \\
					& $\frac\alpha\beta$	&	Ratio of rate of circulating lymphocyte production to turnover rate & \\
							&			&	$3\times10^9$ (Cells L$^{-1}$)						& \cite{aids_factsheet}\\
$-\beta C$					&			&	Lymphocyte turnover								& \\
							&	$\beta$	&	Rate of lymphocyte turnover							& \\
							&			&	$6.3\times10^{-3}$ (Day$^{-1}$)						& \cite{de_pillis_et_al_2009}\\
$-K_C(1-e^{-\delta_CM})C$		&			&	Death of lymphocytes due to chemotherapy toxicity		& \\
							&	$K_C$	&	Rate of lymphocyte depletion from chemotherapy toxicity	& \\
							&			&	$5.7\times10^{-1}$ (Day$^{-1}$)						& \cite{catimel_et_al_1995}\\
							& $\delta_C$	&	Chemotherapy toxicity coefficient						& \\
							&			&	$2\times10^{-1}$ (L mg$^{-1}$)						& \cite{vilar_et_al_2008}\\
\label{tab:lymphos}
\end{longtable}
\begin{flushleft}Descriptions of the biological relevance of each term and parameter and the parameter values in $C(t)$, which tracks the concentration of other lymphocytes.\\
\end{flushleft}
\end{center}

\begin{center}
\begin{longtable}{llp{3in}c}
\caption{\bf{Interleukin-2 Equation Terms and Parameters.}}\\ \hline
&&&\\
\bf{Term}		&	\bf{Parameter}	&	\bf{Description and Value (Units)}			& \bf{Source}\\
\endfirsthead

\multicolumn{4}{c}
{{\bfseries \tablename\ \thetable{} -- continued from previous page}} \\

\bf{Term}		&	\bf{Parameter}	&	\bf{Description and Value (Units)}			& \bf{Source}\\
\endhead

\multicolumn{4}{r}{{\em Table cont'd on next page.}}\\ \hline
\endfoot
\hline\hline
\endlastfoot

$-\mu_II$					&			&	IL-2 turnover										& \\
							&	$\mu_I$	&	Rate of excretion and elimination of IL-2				& \\
							&			&	$11.7427$ (Day$^{-1}$)								& \cite{de_pillis_et_al_2009}\\
$\phi C$						&			&	Production of IL-2 due to naive CD$8^+$T cells and CD$4^+$T cells & \\
							&	$\phi$	&	Rate of IL-2 production from CD$4^+$/naive CD$8^+$T cells & \\
							&			&	$1.788\times10^{-7}$ (IU Cells$^{-1}$Day$^{-1}$)		& \cite{de_pillis_et_al_2009}\\
$\frac{\omega LI}{\zeta+I}$		&			&	Production of IL-2 from activated CD$8^+$T cells			& \\
							& $\omega$	&	Rate of IL-2 production from CD$8^+$T cells			& \\
							&			&	$7.88\times10^{-2}$ (IU Cells$^{-1}$Day$^{-1}$)			& \cite{de_pillis_et_al_2009}\\
							&	$\zeta$	&	Concentration of IL-2 for half-maximal CD$8^+$T-cell IL-2 production & \\
							&			&	$2.5036\times10^3$ (IU L$^{-1}$)						& \cite{de_pillis_et_al_2009}\\
\label{tab:il2}
\end{longtable}
\begin{flushleft}Descriptions of the biological relevance of each term and parameter and the parameter values in $I(t)$, which tracks the concentration of interleukin-2.\\
\end{flushleft}
\end{center}

\begin{center}
\begin{longtable}{llp{3in}c}
\caption{\bf{Medication Equations Terms and Parameters.}} \\ \hline
&&&\\
\bf{Term}	&	\bf{Parameter}		&	\bf{Description and Value (Units)}		& \bf{Source}\\
\endfirsthead

\multicolumn{4}{c}
{{\bfseries \tablename\ \thetable{} -- continued from previous page}} \\

\bf{Term}	&	\bf{Parameter}		&	\bf{Description and Value (Units)}		& \bf{Source}\\
\endhead

\multicolumn{4}{r}{{\em Table cont'd on next page.}}\\ \hline
\endfoot
\hline\hline
\endlastfoot

Chemotherapy					&			&													& \\
$-\gamma M$					&			&	Excretion and elimination of chemotherapy				& \\
							& $\gamma$	&	Rate of excretion and elimination of chemotherapy drug	& \\
							&			&	$4.077\times10^{-1}$ (Day$^{-1}$)					& \cite{catimel_et_al_1995}\\
							&			&													& \\
mAb Therapy					&			&													& \\
$-\eta A$						&			&	Excretion and elimination of mAbs						& \\
							&	$\eta$	&	Rate of mAb turnover and excretion					& \\
							&			&	$1.386\times10^{-1}$ (Day$^{-1}$)$^a$				& \cite{grothey_2006}\\
							&			&	$9.242\times10^{-2}$ (Day$^{-1}$)$^b$				& \cite{grothey_2006}\\
$-\lambda T\frac{A}{h_2+A}$		&			&	Loss of mAbs due to tumor-mAb binding				& \\
							& $\lambda$	&	Rate of mAb-tumor cell complex formation				& \\
							&			& $8.9\times10^{-14}$ (mg Cells$^{-1}$L$^{-1}$Day$^{-1}$)$^a$ & \cite{freeman_et_al_2008}\\
							&			& $8.6\times10^{-14}$ (mg Cells$^{-1}$L$^{-1}$Day$^{-1}$)$^b$ & \cite{freeman_et_al_2008}\\
							&	$h_2$	& Concentration of mAbs for half-maximal EGFR binding		& \\
							&			&	$4.45\times10^{-5}$ (mg L$^{-1}$)$^a$				& \cite{freeman_et_al_2008}\\
							&			&	$4.3\times10^{-5}$ (mg L$^{-1}$)$^b$					& \cite{freeman_et_al_2008}\\
\label{tab:meds}
\end{longtable}
\begin{flushleft}Descriptions of the biological relevance of each term and parameter and the parameter values in $M(t)$ and $A(t)$, which track the concentration of chemotherapy and mAb therapy, respectively.\\
$^a$ For cetuximab.\\
$^b$ For panitumumab.
\end{flushleft}
\end{center}

\clearpage 

\noindent \Large\textbf{Acknowledgment}\\[2mm]

\footnotesize A.E. Radunskaya was partially supported by NSF grant DMS-1016136.  \\[3mm]

\appendix 
\bibliographystyle{plain}
\bibliography{my_refs}
\vfill \eject

\scriptsize\-----------------------------------------------------------------------------------------------------------------------------------------\\\copyright \it 2013  Author1 \& Author2; This is an Open Access article distributed under the terms of the Creative Commons Attribution License
\href{http://creativecommons.org/licenses/by/2.0}{http://creativecommons.org/licenses/by/2.0},  which permits unrestricted use, distribution, and reproduction in any medium,
provided the original work is properly cited.

\begin{flushright}


{\Large \textbf{\\Mathematical Model of Colorectal Cancer with Monoclonal Antibody Treatments }}\\[5mm]
{\large \textbf{L.G. dePillis$^\mathrm{*1}$\footnote{\emph{*Corresponding author: E-mail: depillis@hmc.edu}},
H. Savage$^\mathrm{2}$\\[1mm]
and A.E. Radunskaya$^\mathrm{3}$}}\\[1mm]
$^\mathrm{1}${\footnotesize \it Dept. of Mathematics,\\ Harvey Mudd College \\ Claremont, Califoria, USA}\\ 
$^\mathrm{2}${\footnotesize \it Dept. of Mathematics,\\ Harvey Mudd College \\ Claremont, Califoria, USA}\\ 
$^\mathrm{3}${\footnotesize \it Dept. of Mathematics,\\ Pomona College \\ Claremont, Califoria, USA}
\end{flushright}

\begin{flushleft}\fbox{%
\begin{minipage}{1.3in}
{\slshape \textbf{Research Supplement}\/}
\end{minipage}}
\end{flushleft}

\begin{flushright}\footnotesize \it Received: 10 December 2013\\ 
Accepted: XX December 20XX\\
Online Ready: XX December 20XX
\end{flushright}
\HRule\\[3mm]

\afterpage{
\fancyhead{} \fancyfoot{} 
\fancyfoot[R]{\footnotesize\thepage}
\fancyhead[R]{\scriptsize\it British Journal of Mathematics and Computer Science 
{{X(X), XX--XX}},~2013 \\
 }}

\section{Equilibria and Stability}\label{S1}
The system, without treatment, and for the parameter ranges we use, 
has at least two locally stable equilibrium points.
These points were found numerically.  
Justification for the population sizes and concentrations used
can be found in Appendix \ref{S2}. 
One equilibrium corresponding to the absence of a tumor is $E_0:$

\[T=0,\ N=3.333\times10^8,\ L=2.526\times10^4,\ C=3\times10^9,\ M=0,\]
\[I=48.9273,\ A=0,\]

and a large tumor equilibrium is given by $E_L:$

\[T=4.65928\times10^9,\ N=3.333\times10^8,\ L=5.268\times10^5,\ C=3\times10^9,\]
\[M=0,\ I=1173,\ A=0.\]


A linearization of the system \eqref{eq:tumor} - \eqref{eq:patient} 
about these equilibrium points shows that the eigenvalues of the Jacobian of the 
linear system both at $E_0$ and $E_L$ are strictly negative.  Thus, both $E_0$ and $E_L$ are locally asymptotically stable fixed points.  This is illustrated by the numerical simulations shown in 
Figure~\ref{fig:stability}.

For the \emph{no tumor equilibrium}, $E_0,$ the eigenvalues were found to be:
\[\{-10.98,\ -1.67,\ -0.41,\ -0.14,\ -0.046,\ -0.010,\ -0.0063\},\]
and for the \emph{large tumor equilibrium}, $E_L,$ the eigenvalues were found to be:
\[\{-25.85,\ -9.81,\ -9.43,\ -0.41,\ -0.23,\ -0.0098,\ -0.0063\}.\]

\section{Parameters}\label{Parameters}\label{S2}

	In order to determine parameter values, we searched peer-reviewed literature for \emph{in vitro} and \emph{in vivo} studies of colorectal tumor growth that could provide data for the following cases: no treatment, chemotherapy treatment with irinotecan, mAb treatment with cetuximab, and mAb treatment with panitumumab. Some of the parameters used here are those found by de~Pillis and colleagues~\cite{de_pillis_et_al_2009} and their derivation is not repeated. The description and values for each parameter can also be found in Tables~\ref{tab:tumor}-\ref{tab:meds}.

\subsubsection*{Initial Conditions}\label{InitialConditions}

We determine initial conditions
for both a healthy individual and for 
a colorectal cancer patient who has
previously undergone treatment for the tumor. 
The initial values of $N$, $L$, $C$, and $I$ can be determined for each 
individual
by considering biological arguments for reasonable cell concentrations of patients with a ``strong'' and 
``weak'' immune system. 

	The \emph{no tumor equilibrium} was found by considering a healthy individual 
with no tumor ($T=0$) and receiving no cancer treatments ($M=A=0$). 
Because we are assuming that healthy individuals are in homeostasis, we can set each time derivative equal to zero. $N$ and $C$ were found by assuming a lymphocyte count 
of $3.333\times10^9$ cells per liter of blood, which is within the range for a normal lymphocyte count, and assuming natural killer and CD8$^+$ T cell counts to be 10 percent and $<1$ percent, respectively~\cite{abbas_et_al_2005}. This gives us that $C=3.333\times10^9\times0.9=3\times10^9$ and $N=3.333\times10^9\times0.1=3.333\times10^8$. The values for $L$ and $I$ are 
taken from 
\cite{de_pillis_et_al_2009}, in which $L$ is derived from 
\cite{pittet_et_al_1999} and 
\cite{speiser_et_al_2001}, and $I$ is taken from 
\cite{orditura_et_al_2000} and information provided by~\cite{novartis}. These calculations give us the following values for our \emph{no tumor equilibrium}:
\[T=0,\ N=3.333\times10^8,\ L=2.526\times10^4,\ C=3\times10^9,\ M=0,\]
\[I=48.9273,\ A=0.\]

As discussed section \ref{S1},
these initial conditions correspond to a stable equilibrium state of the system.  As shown
in Figure \ref{fig:stability}, left panel, values starting close to these will be drawn toward this
zero tumor
equilibrium.

	The \emph{large tumor equilibrium} was found by considering a healthy individual who has a large tumor but is not receiving any treatment ($M=A=0$). We again set the time derivatives to zero under the assumption of homeostasis. Under conditions of an untreated tumor, we leave $N$ and $C$ at the same values, but use larger values for $I$ and $L$, since the presence of a tumor increases the production of cytokines~\cite{de_pillis_et_al_2009}. We take the values of $I=1173$ and $L=5.268\times10^5$ from 
\cite{de_pillis_et_al_2009}, in which $L$ is taken from 
\cite{lee_et_al_1999} and Janeway's book on \emph{Immunobiology}~\cite{janeway_et_al_2005}. 
The value of $I$ is from 
\cite{orditura_et_al_2000} and information provided by~\cite{novartis}. 
With these initial values and the parameters that can be directly calculated from available literature, we solve for the size of a large tumor in equilibrium while solving 
for the parameter $p$ in the section on NK cell parameters. Note that the 
resulting value, $T=4.65928\times10^9$, is slightly less than the theoretical carrying 
capacity of $4.66\times10^9$ which we find during the calculation of the parameter $b$ in the section on tumor parameters. 
This is expected, because interactions with the immune system prevent the tumor from reaching its theoretical carrying capacity. These initial values give us the following \emph{large tumor equilibrium}:
\[T=4.65928\times10^9,\ N=3.333\times10^8,\ L=5.268\times10^5,\ C=3\times10^9,\]
\[M=0,\ I=1173,\ A=0.\]
These initial conditions also correspond to a stable equilibrium state of the system,
as discussed in Appendix \ref{S1}.
As illustrated
in Figure \ref{fig:stability}, right panel, 
values starting close to these will be drawn toward this high tumor
equilibrium.

	Since the majority of the individuals we are considering have 
previously undergone various treatments and do not have very strong immune systems, we reduce the initial values for $N,\ L$, and $C$ in our simulations. A normal leukocyte count is $4.5-11\times10^9$ cells/L, and lymphocytes can make up 16-46\% of the total leukocytes~\cite{abbas_et_al_2005}. Thus a normal lymphocyte count is $0.72-5.06\times10^9$ cells/L. We set the initial total lymphocyte count in our simulated individuals to $9.9\times10^8$ cells/L, a value within the normal range for lymphocyte concentration, but close to being low. 
Natural killer cells and activated CD8$^+$T cells interact more directly with the tumor than the other lymphocytes, so we assume that they are deactivated at a slightly higher rate. Thus NK cells constitute a slightly smaller percentage of the total lymphocytes than the normal value of 10\%. We set $N(0)$ to 9\% of total lymphocytes, so $N(0) = .9\times(9.9\times10^8) = 9\times 10^7$. NK cell population was reduced to approximately $\frac 13 - \frac 14$ it's original value, so we reduce $L(0)$ to approximately $\frac 13$ its original value also, and set $L(0)=1.8\times 10^5$. This leaves $C(0)=9\times10^8$. Initial values for $M$ and $A$ are set to zero, but $\frac{dM}{dt}$ and $\frac{dA}{dt}$ will be nonzero at any time $t$ when the patient is receiving treatments. Since the presence of tumor fragments stimulate IL-2 production~\cite{sompayrac_immune}, we leave $I=1173$ as the initial value for $I$. The initial value of $T$ can be varied, and is stated with simulations. These calculations give us the following 
initial conditions for the ``sick'' populations in our model:
\[N(0)=9\times10^7,\ L(0)=1.8\times10^5,\ C(0)=9\times 10^8,\ M(0)=0,\]
\[I(0)=1173,\ A(0)=0.\]
These initial values represent patients who are not in homeostasis, and depending on the initial tumor size, the strength of interactions between the patient's immune system and the tumor, and whether any medication is given, their cell populations can be driven either to the \emph{no tumor equilibrium} or to the \emph{large tumor equilibrium}. Sample conditions for a tumor that is reduced to the \emph{no tumor equilibrium} and for a tumor that grows to the \emph{large tumor equilibrium} 
are found in Figure \ref{fig:stability}.

\subsubsection*{$\frac{dT}{dt}$: The tumor}

For a summary of the terms, parameters, and parameter values, see Table~\ref{tab:tumor}.
	\begin{description}
		\item[$a$] $=2.31\times10^{-1}$ day$^{-1}$, the tumor growth rate, was calculated from the doubling time of colorectal tumors during exponential growth, 
which was found 
in~\cite{corbett_et_al_1975} to be 3 days. We can calculate $a$ from the equation for exponential growth with a half-life of $t=3$ days. So, $2t_0=t_0e^{at}$, giving us 
$a=\frac{\ln(2)}{3}=2.31\times10^{-1}$. This is approximately half of the value for $a$ found by de~Pillis's team for melanoma~\cite{de_pillis_et_al_2009}, but colon tumors are known to have slower growth rates than most of cancers, so this is not an unreasonable value~\cite{bolin_et_al_1983,burnett_et_al_1981}. It is important to note that in \cite{corbett_et_al_1975} 
tumors were grown in non-immunodeficient mice, and our model considers patients who do not have a full-strength immune response, however this was the only study in our literature search that 
provided the doubling time specifically during exponential growth. The growth rate that we calculated also agrees with the initial growth rates found 
in~\cite{leith_et_al_1987}, who grew colon tumors in immunodeficient mice.
		\item[$b$] $=2.146\times10^{-10}$ cells$^{-1}$, is the inverse of the carrying capacity. The theoretical carrying capacity (in volume) of colorectal 
tumors was taken from Leith and colleagues~\cite{leith_et_al_1987}, who collected tumor growth data, fit them to the Gompertz equation, and found the maximum tumor size as $t\rightarrow\infty$. The carrying capacity derived from the Gompertz model has the same biological interpretation as in our model, so we were able to use the results of 
\cite{leith_et_al_1987} to find a value for $b$. Multiple carrying capacities were found from different colorectal tumor lines, with an average of approximately 10,000 mm$^3=10^{13}\ \mu$m$^3$.  This size was then converted to a cell population using 2145 $\mu$m$^3$ as the average tumor cell volume~\cite{chen_ieee_2005}, giving $10^{13}\mu$m$^3/(2145\mu$m$^3/$cell)=$4.66 \times10^9$ cells. Thus, $b=(4.66\times10^9$ cells$)^{-1}=2.146\times10^{-10}$.
		\item[$c$] $=5.156\times10^{-14}$ L cells$^{-1}$day$^{-1}$, the rate of NK-induced tumor death, is set equal to $p$ (see the section on NK cell parameters), as was done 
in~\cite{de_pillis_et_al_2009}, under the assumption that when an NK cell kills a tumor cell, the NK cell also is deactivated. Recent research~\cite{bhat_watzl_2007} suggests that natural killer cells may be able to kill up to six tumor cells before deactivation. However we have not found further confirmation of this and so have chosen to continue using the assumption that NK cells are only able to kill one tumor cell each.
		\item[$D=d\frac{(L/T)^l}{s+(L/T)^l}$] is a patient-specific term that involves three parameters to which we assign four separate values each, in order to reflect a variety of patient-specific states. These parameters are: $d$ (day$^{-1}$), the immune-system strength coefficient; $l$, the immune-system strength scaling coefficient; and $s$ (L), the value of $(\frac LT)^l$ necessary for half-maximal CD$8^+$ T-cell effectiveness against tumor. We base our values for $d,\ l$, and $s$ on the values of $d\in\{1.88,2.34\},\ l\in\{1.81,2.09\}$, and $s\in\{3.5\times10^{-2},3.8\times10^{-3}\}$ used 
in~\cite{de_pillis_et_al_2009}, and slightly weaken the patient immune system (represented by lowering $d$ and $l$ and raising $s$) to represent 
individuals who
are not in 
good health 
from having gone through multiple cancer treatments. We use $d\in\{1.3,1.6,1.9,2.1\},\ l\in\{1.1,1.4,1.7,2.0\}$, and $s\in\{4\times10^{-3},7\times10^{-3},9\times10^{-3},3\times10^{-2}\}$, 
which results in sixty-four different 
individual immune profiles
over which we can run simulations to represent clinical trials.
		\item[$\xi$] $=6.5\times10^{-10}$ L cells$^{-1}$day$^{-1}$ for cetuximab, and $=0$ for panitumumab, is the rate of NK-induced tumor death through ADCC. The value for cetuximab was set to match the expected increase in NK cell activity found by Kurai and colleagues~\cite{kurai_et_al_2007}. Kurai's team varied concentrations of tumor cells and NK cells, left them for 4 hours with and without $0.25\mu$g/mL cetuximab, and measured the resulting NK activity. They measured the activity at much higher concentrations of NK cells than are present in the body, but based on their results we approximated that at the ratio of one NK cell to ten tumor cells, NK activity is increased by 10 percent. We found an appropriate value for $\xi$ by running simulations with varying values of $\xi$ and 
simulating their experimental conditions: $t=4$ hours, $T_0=10^9$, $N_0=\frac 14\times T_0=2.5\times10^8$, and an initial treatment of 0.25 mg/L cetuximab over 15 minutes. The other immune system components, as well as natural growth and decay, were not included. A value of $\xi=6.5\times10^{-10}$ was found to give the desired 10 percent decrease in NK cells in this experiment, which we use as a proxy for an increase in NK activity of 10 percent. Panitumumab is unable to activate the ADCC pathway, so $\xi$ is set to zero in that case~\cite{grothey_2006}.
		\item[$h_1$] $=1.25\times10^{-6}$ mg L$^{-1}$ for cetuximab, and $0$ for panitumumab, is the concentration of mAbs necessary for a half-maximal increase in ADCC. The ADCC activity level indicated by $\xi$ is reached when the cetuximab concentration is above $0.25\mu$g/mL, and so $h_1$ was set to $.5\times0.25\mu$g/mL$=1.25\times10^{-6}$ mg/L. Cetuximab levels in the body are usually above this threshold during treatment, and we have chosen to use a sigmoid function to capture this threshold. Although we do 
not have evidence to support that ADCC activity increases according to a saturation function, this model captures two important characteristics: that the threshold concentration for maximal ADCC activity is much lower than the normal cetuximab dose, and that the ADCC activity level approaches zero as mAb concentration approaches zero. We chose $h_1$ so that, when the cetuximab concentration is half of the threshold value, the term $\frac A{h_1+A}$ equals one half, resulting in half-maximal ADCC activity. Because panitumumab does not play a role in ADCC, panitumumab does not have an $h_1$ ($h_1=0$).
		\item[$K_T$] $=8.1\times10^{-1}X$ day$^{-1}$ is the rate of chemotherapy-induced tumor death, where $X$ is a random variable with \emph{probability density function} 
$p(x) = \frac{1}{3}(1-x)^{-2/3}, 0 \leq x < 1.$
We chose this distribution since it is supported on $[0,1],$ 
has a high probability of being close to one, and
a mean of $E[X]= 0.75.$
Therefore, $K_T\in[0,8.1\times10^{-1}]$, and has a mean value of $K_T=6.075\times10^{-1}$. 
Note that, in the clinical trial simulations, each patient
is assigned a value for $K_T$, but a different $K_T$ is randomly generated for each patient according to the distribution given above. 

	The maximal value of $K_T$ was calculated from \emph{in vitro} data collected by Vilar and colleagues~\cite{vilar_et_al_2008} on irinotecan concentration and growth reduction of various colon cancer cell strains. We chose to use values from the HT-29 cell line, in accordance with much of the literature that we reviewed.  We estimated five coordinates from data in \cite{vilar_et_al_2008}, which gave irinotecan concentration (in mol/L) versus growth of tumor cells, as a percentage of tumor cell growth with no irinotecan. Since the reported data was from an  \emph{in vitro} study run over the course of only a few days, we set all but tumor size and chemotherapy concentration to zero and assumed that the natural cell death was zero. We also assumed that chemotherapy concentration would be held constant, so $\frac{dM}{dt}=0$. Thus the differential equation for the tumor population becomes $\frac{dT}{dt}=-K_T(1-e^{-\delta_T M}) T$,  with solution:  $T=T_0 e^{-K_T(1-e^{-\delta_T M} ) t}$. We converted the irinotecan concentration at each point to units of mg/L using 677 g/mol as the molecular weight of irinotecan~\cite{tsuruo_et_al_1988}. We then used tumor sizes and chemotherapy concentrations from each data point reported in \cite{vilar_et_al_2008} to write five equations with $\delta_T$ and $K_T$ as unknowns. Since the system is overdetermined (five equations, two unknowns), we chose values for $\delta_T$ and $K_T$ that produced a reasonable fit. We found $\delta_T=0.2$ and $K_T\approx0.85$.

	$K_T$ was then separately confirmed by running multiple simulations with our set of patient-specific parameter values to look for a tumor response rate of approximately 15-20\% after 6 weeks of treatment. The reported \emph{response rate} (the percentage of patients whose tumor was not larger after treatment) for irinotecan is around 30 percent, however patients receiving mAb treatment have usually already received a variety of chemotherapy treatments and did not respond strongly to them, so we aimed for a response rate lower than this~\cite{cancer_tx_book}. These simulations confirmed that a value of $K_T=0.81$ gives an average response rate of approximately 19 percent.
		\item[$K_{AT}$] $=4\times10^{-4}$ L mg$^{-1}$day$^{-1}$ for both cetuximab and panitumumab, is the additional chemotherapy-induced tumor death due to mAb-tumor interactions. Even with $K_{AT}$ set to zero, our simulation response rates are much higher than those reported in clinical trials, however, it is known that mAb therapy can help to increase chemotherapy re\-spon\-ses in tumors, and even restore partial response in chemo\-ther\-apy-re\-frac\-tory tumors, so we have chosen to give $K_{AT}$ a non-zero value of $K_{AT}=4\times10^{-4}$~\cite{cancer_tx_book}. At maximal mAb concentrations, which are on the order of $10^2$ mg, this results in an increase in chemotherapy activity of approximately $\frac{4\times10^{-4}\times10^2}{8.1\times10^{-1}}\approx.05=5\%$.
		\item[$\delta_T$] $=2\times10^{-1}$ L mg$^{-1}$, the medicine efficacy coefficient, was found as part of the calculation for $K_T$.
		\item[$\psi$] $=2.28\times10^{-2}Y$ L mg$^{-1}$day$^{-1}$ for cetuximab and $3.125\times10^{-2}y$ L mg$^{-1}$day$^{-1}$ for panitumumab is the rate of mAb-induced tumor death, where $Y$ is a random variable with probability density function $p(y)=\frac{1}{3}(1-y)^{-2/3}, 0 \leq y < 1.$
Therefore, $\psi\in[0,2.28\times10^{-2}]$ for cetuximab, with a mean value of $1.71\times10^{-2}$, and $\psi\in[0,2.58\times10^{-2}]$ for panitumumab, with a mean value of $1.94\times10^{-2}$. As with $K_T$, multiplying the maximum value for $\psi$ by a random variable between zero and one allows us to represent that each tumor has a different response to treatments. Each patient (each simulation) has one constant value for $\psi$, but a different $\psi$ is randomly generated for every patient.

The maximum value of $\psi$ was found by running simulations of mAb therapy over a range of possible values for $\psi,$ using the full set of
patient-specific parameters.  The values of $\psi$ we chose yielded a 10\% response rate for cetuximab at four weeks, and a 12.2\% response rate for panitumumab at six weeks.  These response rates reflect those reported
in \cite{cancer_tx_book}.
	\end{description}

\subsubsection*{$\frac{dN}{dt}$: Natural killer cells}

For a summary of the terms, parameters, and parameter values, see Table~\ref{tab:nk_cells}.
	\begin{description}
		\item[$\frac ef$] $=\frac 19$, the ratio of the NK cell synthesis rate to the turnover rate, is found using the same method as 
was used in~\cite{de_pillis_et_al_2009}. The value for $\frac ef$ is found by assuming the \emph{no tumor equilibrium} 
and thus setting $T=0$ and setting Equation~\ref{eq:natural_killers} to zero. We then ignore the term $\frac{p_NNI}{g_N+I}$, which has only a very small effect on NK proliferation. This gives us $f(\frac ef C-N)=0$, and so $\frac ef = \frac NC$. As in the equilibrium calculations, NK cells make up approximately 10 percent of all lymphocytes, and T cell count is negligible, giving us $\frac{10\%}{90\%}$, or $\frac 19$~\cite{abbas_et_al_2005}.
		\item[$f$] $=1\times10^{-2}$ day$^{-1}$, the rate of NK cell turnover, is based on the value of $f=1.25\times10^{-2}$ found by de~Pillis and colleagues~\cite{de_pillis_et_al_2009}. We lowered the value slightly to agree with our assumption of a patient with a weakened immune system whose body may not be able to produce new cells as quickly as normal healthy individual.
		\item[$g_N$] $=2.5036\times10^5$ IU L$^{-1}$, the concentration of IL-2 needed for half-maxi\-mal NK cell proliferation, is unchanged from the value found by 
in~\cite{de_pillis_et_al_2009}.
		\item[$p_N$] $=5.13\times10^{-2}$ day$^{-1}$, the rate of IL-2 induced NK cell proliferation, is calculated using the same method as 
in~\cite{de_pillis_et_al_2009}. They use data from 
\cite{meropol_et_al_1998} to find that $5.0073\times10^4$ IU stimulates NK cells to reach a count of $2.3\times10^9$ cells, and so using these as $I$ and $N$ respectively and assuming $T=0$, we then set Equation~\ref{eq:natural_killers} equal to zero and solve for $p_N$: \[p_N=\frac{f(N-\frac ef C)(g_N+I)}{NI}.\] Using $C=3\times10^9$ from our \emph{no tumor equilibrium} and the previously calculated values for $e,\ f$, and $g_N$, we find that $p_N=5.13\times10^{-2}$.
		\item[$p$] $=5.156\times10^{-14}$ L cells$^{-1}$ day$^{-1}$, the rate of NK cell death due to tumor interaction, is calculated using the same method as 
in~\cite{de_pillis_et_al_2009}. We consider the \emph{large tumor equilibrium} with no medication, assume (as explained in the calculation for $c$) that $p=c$, and can thus set Equations~\ref{eq:tumor} and~\ref{eq:natural_killers} equal to zero and to solve for $T$ and $p$:
		\begin{align*}
			p & =\frac{{\frac{p_N N I}{g_N+I}+eC-fN}}{NT} \text{ and}\\
			0 & = aT(1-bT)-cNT-DT\\
			{}& = aT(1-bT)-pNT-DT.
		\end{align*}
		We were then able to use the values for $p_N,\ g_N,\ e,\ f,\ a,\ b$, the equation for $D$ with the moderate patient-specific parameter values of $d=1.9$, $l=1.6$, and $s=7\times10^{-3}$, and the state values for the immune system populations from the \emph{large tumor equilibrium} to find that $T=4.65928\times10^9$ in the \emph{large tumor equilibrium} and $p=5.156\times10^{-14}$.
		\item[$p_A$] $=6.5\times10^{-10}$ L cells$^{-1}$day$^{-1}$ for cetuximab and $0$ for panitumumab is the rate of NK cell death due to interactions with mAb-tumor complexes. We set $p_A=\xi$, under the approximation used for the calculation of parameter $c$ that for each tumor cell killed through ADCC, one NK cell also dies.
		\item[$K_N$] $=9.048\times10^{-1}$ day$^{-1}$, the rate of NK depletion from chemotherapy toxicity, is calculated using the same method as 
in~\cite{de_pillis_et_al_2009}, by linearly scaling $K_C$ by the ratio of cell metabolic rates. That is, \[K_N=\frac f\beta K_C.\]
		\item[$\delta_N$] $=2\times10^{-1}$ L mg$^{-1}$, the chemotherapy toxicity coefficient, is assumed to equal $\delta_T$. The drug has a different efficacy ($K$) for each cell type, but we assume that a similar concentration of irinotecan is needed to affect each cell, regardless of cell type~\cite{de_pillis_et_al_2009}.
	\end{description}

\subsubsection*{$\frac{dL}{dt}$: CD8$^+$T cells}

For a summary of the terms, parameters, and parameter values, see Table~\ref{tab:t_cells}.
	\begin{description}
		\item[$m$] $=5\times10^{-3}$ day$^{-1}$, the rate of activated CD8$^+$T-cell turnover, is based on the value of $m=9\times10^{-3}$ found by de~Pillis and colleagues~\cite{de_pillis_et_al_2009}. We lowered the value slightly to agree with our assumption of a patient with a weakened immune system, whose body may not be able to produce new cells as quickly as normal healthy individual.
		\item[$\theta$] $=2.5036\times10^{-3}$ IU L$^{-1}$, the concentration of IL-2 to halve CD8$^+$T-cell turnover, is unchanged from de~Pillis and colleagues~\cite{de_pillis_et_al_2009}.
		\item[$q$] $=5.156\times10^{-17}$ cells$^{-1}$day$^{-1}$, the rate of CD8$^+$T-cell death due to tumor interaction, is set equal to $p\times10^{-3}$ because, as de~Pillis and colleagues~\cite{de_pillis_et_al_2009} point out, we expect $q$ to be approximately three orders of magnitude less than $p$ since $L$ is approximately three orders of magnitude less than $N$.
		\item[$r_1$] $=5.156\times10^{-12}$ cells$^{-1}$day$^{-1}$, the rate of NK-lysed tumor cell debris activation of CD8$^+$T cells, is calculated using the same method as 
in~\cite{de_pillis_et_al_2009}. We set $r_1=100\times c$, based on the approximation that a lysed tumor cell can stimulated 10-300 T cells per day~\cite{de_pillis_et_al_2009}.
		\item[$r_2$] $=1\times10^{-15}$ cells$^{-1}$day$^{-1}$, the rate of CD8$^+$T-cell production from circulating lymphocytes, is based on the value of $r_2=5.8467\times10^{-13}$ found by de~Pillis and colleagues~\cite{de_pillis_et_al_2009}. 
We reduced it from 
the value in~\cite{de_pillis_et_al_2009}
to reflect that a weakened immune system may not be able to produce activated CD8$^+$T cells as effectively.
		\item[$p_I$] $=2.4036$ day$^{-1}$, the rate of IL-2 induced CD8$^+$T-cell activation, was found using the same method as 
in~\cite{de_pillis_et_al_2009}. A system of equations was created by considering the \emph{no tumor equilibrium} and the \emph{large tumor equilibrium}. Setting Equation~\ref{eq:killer_ts} to zero and using these two sets of initial values for $T,\ N,\ L,\ C$, and $I$, we can obtain two equations each with $p_I$ and $u$ as unknowns, and thus solve for the $p_I$ and $u$ necessary to make satisfy the equilibrium conditions.
		\item[$g_I$] $=2.5036\times10^3$ IU L$^{-1}$, the concentration of IL-2 necessary for half-maximal CD8$^+$T-cell activation, is unchanged from the value found 
in~\cite{de_pillis_et_al_2009}.
		\item[$u$] $=3.1718\times10^{-14}$ L$^2$ cells$^{-2}$day$^{-1}$, the CD8$^+$T-cell self-limitation feedback coefficient, is obtained from the system of equations used to calculated $p_I$.
		\item[$\kappa$] $=2.5036\times10^3$ IU L$^{-1}$, the concentration of IL-2 to halve the magnitude of CD8$^+$T-cell self-regulation, is unchanged from the value found 
in~\cite{de_pillis_et_al_2009}.
		\item[$j$] $=1.245\times10^{-4}$ day$^{-1}$, the rate of CD8$^+$T-cell lysed tumor cell debris activation of CD8$^+$T cells, is based on the value of $1.245\times10^{-2}$ found by de~Pillis and colleagues~\cite{de_pillis_et_al_2009}, and was decreased to indicate that the weak immune system may not be able to activate CD8 cells as effectively.
		\item[$k$] $=2.019\times10^7$ cells, the tumor size for half-maximal CD8$^+$T-cell lysed tumor debris CD8$^+$T cell activation, is unchanged from the value found 
in~\cite{de_pillis_et_al_2009}.
		\item[$K_L$] $=4.524\times10^{-1}$ day$^{-1}$, the rate of CD8$^+$T-cell depletion from chemotherapy toxicity, is found in the same way as we found $K_N$. We calculated it using the same method as 
was used in~\cite{de_pillis_et_al_2009}, by linearly scaling $K_C$. That is, \[K_L=\frac m\beta K_C.\]
		\item[$\delta_L$] $=2\times10^{-1}$ L mg$^{-1}$, the chemotherapy toxicity coefficient, is found in the same way as $\delta_N$, with the assumption that it is equal to $\delta_T$~\cite{de_pillis_et_al_2009}.
	\end{description}

\subsubsection*{$\frac{dC}{dt}$: Lymphocytes}

For a summary of the terms, parameters, and parameter values, see Table~\ref{tab:lymphos}.
	\begin{description}
		\item[$\frac\alpha\beta$] $=3\times10^9$ cells L$^{-1}$, the ratio of the rate of circulating lymphocyte production to turnover rate, is taken from considering the steady state assumption of $\frac{dC}{dt}=0$ in a healthy, tumor free individual. Considering Equation~\ref{eq:lymphocytes} with $M=0$, we find that $\frac\alpha\beta=C$, where $C=3\times10^9$ refers to the equilibrium value of $C$ in the \emph{no tumor equilibrium}.
		\item[$\beta$] $=6.3\times10^{-3}$ day$^{-1}$, the rate of lymphocyte turnover, is unchanged from the value found 
in~\cite{de_pillis_et_al_2009}.
		\item[$K_C$] $=5.7\times10^{-1}$ day$^{-1}$, the rate of lymphocyte depletion from chemotherapy toxicity, was calculated to achieve the results given by Catimel and colleagues~\cite{catimel_et_al_1995} on the number of patients with leukopenia after irinotecan treatments. Catimel's team found that when 100 mg/m$^2$ was given to patients daily for three days, three out of eleven patients had leukopenia, and when 115 mg/m$^2$ was given daily for three days, four out of ten patients had leukopenia. A patient is considered to have leukopenia when the leukocyte count drops below $1.9\times10^9$~\cite{welink_et_al_1999}, and 
as discussed in 
section \ref{S2},
lymphocytes can comprise up to 46\% of leukocytes, so the highest possible lymphocyte count in a patient with leukopenia is 46\% of $1.9\times10^9$, or $8.74\times10^8$ cells. We assume that the lymphocyte count for all patients drop equally, and so those who begin initially with a lower lymphocyte count become leukopenic, and those who begin with a higher lymphocyte count will have a reduced cell count, but remain within the normal range. So, the lowest three elevenths of patients will have lymphocyte levels below $1.904\times10^9$ cells, and the lowest four tenths of patients will have lymphocyte levels below $2.456\times10^9$ cells. We ran simulations considering only lymphocyte counts, with irinotecan delivered once daily over 1.5 hours for a total of 3 days, and found a value for $K_C$ that made an initial lymphocyte count of $1.904\times10^9$ drop to approximately $8.74\times10^8$ with a 100 mg/m$^2$ dose and an initial lymphocyte count of $2.456\times10^9$ drop to approximately $8.74\times10^8$ with a 115 mg/m$^2$ dose. The two doses resulted in $K_C$ values of 0.52 and 0.63 respectively, so these were averaged to find $K_C=.57$.
		\item[$\delta_C$] $=2\times10^{-1}$ L mg$^{-1}$, the chemotherapy toxicity coefficient, is found in the same way as we found $\delta_L$, with the assumption that it is equal to $\delta_T$~\cite{de_pillis_et_al_2009}.
	\end{description}

%

\subsubsection*{$\frac{dI}{dt}$: Interleukin}

For a summary of the terms, parameters, and parameter values, see Table~\ref{tab:il2}.
	\begin{description}
		\item[$\mu_I$] $=11.7427$ day$^{-1}$, the rate of excretion and elimination of IL-2, is unchanged from the value found 
in~\cite{de_pillis_et_al_2009}.
		\item[$\omega$] $=7.88\times10^{-2}$ IU cells$^{-1}$day$^{-1}$, the rate of IL-2 production from CD8$^+$T cells, is calculated using the same method as 
was used in~\cite{de_pillis_et_al_2009}, from the no tumor and large tumor equilibria. $\frac{dI}{dt}$ is set to zero, and the known parameters and initial values are used to find two equations with the two unknowns $\omega$ and $\phi$. We then solve for these two parameters.
		\item[$\phi$] $=1.788\times10^{-7}$ IU cells$^{-1}$day$^{-1}$, the rate of IL-2 production from CD4$^+$ and naive CD8$^+$T-cell IL-2 production, is found as part of the system of equations solving for $\omega$.
		\item[$\zeta$] $=2.5036\times10^3$ IU L$^{-1}$, the concentration of IL-2 for half-maximal CD8$^+$T-cell IL-2 production, is unchanged from the value found 
in~\cite{de_pillis_et_al_2009}.
	\end{description}

\subsubsection*{$\frac{dM}{dt}$: Irinotecan chemotherapy treatment}

For a summary of the terms, parameters, and parameter values, see Table~\ref{tab:meds}.
	\begin{description}
	\item[$\gamma$] $=4.077\times10^{-1}$ day$^{-1}$, the rate of excretion and elimination of chemotherapy drug, is calculated using the assumption of exponential decay from $\frac{\ln(2)}{t_{1/2}}$, where $t_{1/2}$ is the half-life of SN-38, the active form of irinotecan, in tissue. The half-life of irinotecan in rat tissue is 7.2 hours, the half life of irinotecan in rat plasma is 1.8 hours, the half life of irinotecan in human plasma is 8.3 hours, and the half life of SN-38 in human plasma is 10.2 hours~\cite{noble_et_al_2006,catimel_et_al_1995}. So, we assume that the ratio of the irinotecan half life in rat tissue/rat plasma equals the ratio of irinotecan half life in human tissue/human plasma to get that the half life of irinotecan in human tissue is $7.2\times8.3/1.8=33.2$ hours. We also assume that the ratio of irinotecan half life in human tissue/plasma equals the SN-38 half life in human tissue/plasma, which gives us that the half life of SN-38 in human tissue is $33.2\times10.2/8.3=40.8$ hours. Thus $\gamma=\frac{\ln(2)}{40.8/24}=4.077\times10^{-1}$.
	\end{description}

\subsubsection*{$\frac{dA}{dt}$: Cetuximab and panitumumab monoclonal antibody treatment}

For a summary of the terms, parameters, and parameter values, see Table~\ref{tab:meds}.
	\begin{description}
		\item[$\eta$] $=1.386\times10^{-1}$ day$^{-1}$ for cetuximab and $9.242\times10^{-2}$ day$^{-1}$ for panitumumab is the rate of mAb turnover and excretion. The parameter $\eta$ is calculated using the assumption of exponential decay from $\frac{\ln(2)}{t_{1/2}}$, where $t_{1/2}$ is the half-life in tissue of each mAb. For cetuximab, the half life in tissue is 5 days, so $\eta=\frac{\ln(2)}{5}=0.139$~\cite{grothey_2006}. For panitumumab, the half life in tissue is 7.5 days, so $\eta=\frac{\ln(2)}{7.5}=0.092$~\cite{grothey_2006}.
		\item[$\lambda$] $=8.9\times10^{-14}$ mg cells$^{-1}$L$^{-1}$day$^{-1}$ for cetuximab and $8.6\times10^{-14}$ mg cells$^{-1}$L$^{-1}$ day$^{-1}$ for panitumumab is the rate of mAb/tumor-cell complex formation. Average cells have around 20,000 EGFRs~\cite{aacr}. The binding affinity of cetuximab is 400 pM (picomolar, which measures the ratio of the concentration of unbound molecules to the concentration of bound molecules) and for panitumumab it is 50 pM~\cite{freeman_et_al_2008}. We first consider cetuximab, which has a molecular weight of 152 kD=$152\times10^6$ mg/mol~\cite{cetux_mw}. A binding affinity of 400 pM means that 400 pM = [cetuximab][EGFRs]/[cetuximab-EGFR complexes]. We first need to find the number of cetuximab-EGFR complexes per cell:
		\[
			\frac{400\text{ pmol}}{1\text{L}}\times\frac{1\text{ mol}}{10^{12}\text{ pmol}}=4\times10^{-10}\text{ mol/L}.
		\]
So, for each free cetuximab molecule and EGFR, there are $2.5\times10^9$ cetuximab-EGFR complexes. So, out of the 20,000 EGFRs per cell, we expect $<1$ ($8\times10^-6$) EGFR per cell to be free. Thus we will assume that all EGFRs are filled. We can convert this back into concentration of cetuximab lost per tumor cell:
		\begin{align*}
			20,000\text{ mAbs}\times\frac{1\text{ mol}}{6\times10^{23}\text{ mAbs}}&\times\frac{152\times10^6\text{ mg}}{1\text{ mol}}\times\frac{1}{57\text{L}}\\
			&=8.9\times10^{-14}\text{ mg/L}.
		\end{align*}
Thus, for cetuximab, $\lambda=8.9\times10^{-14}$. For panitumumab, we perform a similar computation, using instead panitumumab's binding affinity and its molecular weight of 147 kD=$147\times10^6$ mg/mol~\cite{panit_mw}. We first find the number of panitumumab-EGFR complexes per cell:
		\[
			\frac{50\text{ pmol}}{1\text{L}}\times\frac{1\text{ mol}}{10^{12}\text{ pmol}}=5\times10^{-11}\text{ mg/L}.
		\]
So, for each free panitumumab molecule and EGFR, there are $2\times10^10$ panitumumab-EGFR complexes, and we again assume that all EGFRs are filled. We can convert this back into mg of panitumumab lost per tumor cell:
		\begin{align*}
			20,000\text{ mAbs}\times\frac{1\text{ mol}}{6\times10^{23}\text{ mAbs}}&\times\frac{147\times10^6\text{ mg}}{1\text{ mol}}\times\frac{1}{57\text{L}}\\
			&=8.6\times10^{-14}\text{ mg/L}.
		\end{align*}
Thus for panitumumab, $\lambda=8.6\times10^{-14}$.
		\item[$h_2$] $=4.45\times10^{-5}$ mg L$^{-1}$ for cetuximab and $4.3\times10^{-5}$ mg L$^{-1}$ for panitumumab is the concentration of mAbs for half-maximal EGFR binding. We first consider cetuximab, and use $10^9$ as the number of tumor cells and 57 L as the volume of an average person~\cite{de_pillis_et_al_2009}. Assuming that 20,000 cetuximab molecules bind to each tumor cell, we want to find the number of mg/L at which the EGFRs are saturated:
		\begin{align*}
			\frac{20,000\text{ mAb}}{1\text{ cell}}\times10^9\text{ tumor cells}&\times\frac{1\text{ mol}}{6\times10^{23}\text{ mAb}}  \times\\
			\frac{152  \times  10^6\text{ mg}}{1\text{ mol}}\times\frac{1}{57\text{ L}} & = 8.9\times10^{-5}\text{ mg/L}.
		\end{align*}
So, we set $h_2=0.5\times8.9\times10^{-5}=4.45\times10^{-5}$ mg/L for cetuximab. We perform similar computation for panitumumab:
		\begin{align*}
			\frac{20,000\text{ mAb}}{1\text{ cell}}\times10^9\text{ tumor cells}&\times\frac{1\text{ mol}}{6\times10^{23}\text{ mAb}}  \times\\
			\frac{147  \times  10^6\text{ mg}}{1\text{ mol}}\times\frac{1}{57\text{ L}} & = 8.6\times10^{-5}\text{ mg/L}.
		\end{align*}
Thus we set $h_2=0.5\times8.6\times10^{-5}=4.3\times10^{-5}$ mg/L for panitumumab.
	\end{description}

\subsection*{Treatments}

In this section we show the calculations performed to find the treatment functions ($v_M$ and $v_A$) for the most common treatment schedules. 
We also used other dosing schedules in 
section \ref{J},
but the methods for computing them were the same as those we show here.
Unless otherwise noted, the treatment regimens have been adapted from De Vita's book titled \emph{Cancer: Principles and Practice of Oncology}~\cite{cancer_tx_book}.

\subsubsection*{Irinotecan Treatments}

	$v_M$ (mg/L/day) has been changed to fit a common treatment regimen for irinotecan. 
A 125 mg/m$^2$ dose of irinotecan is usually given over 90 minutes once weekly,
and we give it in our simulations for 4 weeks. We assume 1.73 m$^2$ to be the average surface area of an adult~\cite{ratain_1998}. Because the medication quickly leaves the blood stream, we use 59.71 L, the average volume of an adult, as the volume over which the medication is spread~\cite{de_pillis_et_al_2009}. So, we would like each dose to infuse
	\[125 \frac{\text{ mg}}{\text{ m}^2}\times 1.73 \text{ m}^2 \times (59.71 \text{ L}^{-1}) = 3.6217 \text{ mg/L},\]
	and because it is given over 90 minutes=0.0625 days, we want to set
	\begin{align*}
	v_M(t)=
		\begin{cases}
		57.947 \text{ mg/L/day}	& \text{if treatment was given at time } (t-2/24),
		\\
		0												&	\text{otherwise}.
		\end{cases}
	\end{align*}%

	We check for treatments at time $(t-2/24)$ because irinotecan needs to be converted by the body to its active form, SN-38, and SN-38 levels reach their peak two hours after irinotecan levels~\cite{cancer_tx_book}.

\subsubsection*{Cetuximab Treatments}
	
	For cetuximab, a loading dose of 400 mg/m$^2$ is usually given over two hours, followed by a weekly 250 mg/m$^2$ dose over 60 minutes.
Cetuximab is given on a six-week periodic schedule, during which it is given weekly for the first four weeks, then not given for two weeks. We assume the same surface area and volume as in the previous section. For the loading dose, we would like to infuse
\[400 \frac{\text{ mg}}{\text{ m}^2}\times 1.73 \text{ m}^2 \times (59.71 \text{ L}^{-1} = 11.59 \text{ mg/L},\]
and because it is given over two hours=0.0833 days, we want to set
\[v_A(t)=\frac{11.59 \text{ mg/L}}{0.0833 \text{ days}}=139.072\]
for the first two hours of the simulations. For the regular weekly treatments, we would like to infuse
\[250 \frac{\text{ mg}}{\text{ m}^2}\times 1.73 \text{ m}^2 \times (59.71 \text{ L}^{-1} = 7.243 \text{ mg/L},\]
and because it is given over 60 minutes=0.04167 days, we want to set
\[v_A(t)=\frac{7.243 \text{ mg/L}}{0.04167 \text{ days}}=173.840.\]
Thus, at any time $t$,
	\begin{align*}
	v_A(t)=
		\begin{cases}
		139.072 \text{ mg/L/day}	& \text{if } t\in (0,\frac 2{24})
		\\
		173.840	\text{ mg/L/day}	&	\text{if treatment was given at time } t\ \&\ t\geq\frac 2{24}
		\\
		0													& \text{if treatment was not given at time } t.
		\end{cases}
	\end{align*}

\subsubsection*{Panitumumab Treatments}
	
	The value of $v_A$ for panitumumab was found in the same way, except that 
panitumumab does not require a loading dose. 
We assume a treatment regimen of 6 mg/kg every two weeks, for a total of three treatments. We assume that the medication is given over 60 minutes, and that an average adult weighs of 70 kg~\cite{lewis_merck_2009}. This gives us
	\begin{align*}
	v_A(t)=
		\begin{cases}
		168.816 \text{ mg/L/day}	& \text{if treatment was given at time } t
		\\
		0													&	\text{if treatment was not given at time } t.
		\end{cases}
	\end{align*}



\begin{center}
\begin{longtable}{lll | lll | lll | c}
\caption{\bf{Treatment Response Rates: Simulations compared to Published Results.}} \\ 
\hline
 	                &  			&  		& \multicolumn{3}{c|}{Our Results} 			& \multicolumn{3}{c|}{Published Results}		&        \\ 
Medication$^a$		& Dose 			& Freq. 	& N			& NR			& R			& N		& NR 		& R	& Source \\ \hline

\endfirsthead

\multicolumn{10}{c}%
{{\bfseries \tablename\ \thetable{} -- continued from previous page}} \\

 	                &  			&  		& \multicolumn{3}{c|}{Our Results} 			& \multicolumn{3}{c|}{Published Results}		&        \\ 
Medication$^a$		& Dose 			& Freq. 	& N			& NR			& R			& N		& NR 		& R	& Source \\ \hline

\endhead

\multicolumn{10}{r}{{\em Table cont'd on next page.}}\\ \hline
\endfoot
\hline\hline
\endlastfoot

Irinotecan			& 125 mg/m$^2$	& q1w	& 320		& 81.3\%		& 18.7\%		& NP		& 70\%		& 30\% $^b$	& \cite{cancer_tx_book}\\
 & & & & & & & & & \\
Cmab			& 400 mg/m$^2$	& load \&	& 320		& 90.0\%		& 10.0\%		& NP		& 89-91\%	& 9-11\%		& \cite{cancer_tx_book}\\		
{}				& 250 mg/m$^2$	& q1w	& {}			& {}			& {}			& 346	& 88.4\%		& 11.6\%		& \cite{lenz_2007}\\
{}				& {}				& {}		& {}			& {}			& {}			& 111	& 89.2\%		& 10.8\%		& \cite{cunningham_et_al_2004}\\
 & & & & & & & & & \\
Pmab			& 6 mg/kg		& q2w	& 320		& 87.8\%		& 12.2\%		& NP		& 87\%		& 13\%		& \cite{cancer_tx_book}\\
{}				& {}				& {}		& {}			& {}			& {}			& 231	& 90\%		& 10\%		& \cite{gravalos_et_al_2009}\\
 & & & & & & & & & \\
Irinotecan			& 125 mg/m$^2$	& q1w,	& 320 $^d$	& 3.9\%		& 96.1\%		& NP		& 77.1-5\%	& 22.5-9\%	& \cite{cancer_tx_book}\\
and Cmab		& 400 mg/m$^2$	& load \& & 320		& 83.1\%		& 16.9\%		& NP		& 77\%		& 23\%		& \cite{grothey_2006}\\		
{}				& 250 mg/m$^2$	& q1w	& {}			& {}			& {}			& NP		& 77.1\%		& 22.9\%		& \cite{cunningham_et_al_2004} $^c$\\
 & & & & & & & & & \\
Irinotecan			& 125 mg/m$^2$	& q1w,	& 320 $^d$	& 14.4\%		& 85.6\%		& 34		& 80\%		& 20\%		& \cite{gravalos_et_al_2009} $^c$\\
and Pmab			& 6 mg/kg		& q2w	& 320		& 82.9\%		& 17.1\%		& {}		& {}			& {}			& {}\\ \hline
%

\label{tab:response_rates_common}
\end{longtable}
\begin{flushleft}Response rates for common treatment schedules from clinical trials and from our simulations.\\
$^a$ Abbreviations: Pmab=panitumumab; Cmab=cetuximab; q1w=every week; q2w=every two weeks; q2w=every three weeks; load=loading dose; N=number of patients; NR=no response; R=response, NP=not provided.\\
$^b$ Most response rates for irinotecan found in the literature are for irinotecan as a first-line treatment, including this one. However, patients receiving mAb therapy are usually receiving it because they did not respond well to chemotherapy~\cite{cunningham_et_al_2004}.\\
$^c$ Irinotecan dosing schedule was varied during the study.\\
$^d$ The first response rates (RRs) are measured 7 days after completion of first treatment. The second RRs for each are measured 4 weeks after treatments have ended.
\end{flushleft}
\end{center}

\vfill

\begin{center}
\begin{longtable}{lll | llll}
\caption{\bf{Simulation Response Rates: Combination Therapies with Hypothetical Dosing Schedules}} \\
\hline

 				& 					& 				& \multicolumn{4}{c}{Our Results$^a$ }\\ 
Medication$^b$ 				& Dose 				& Frequency$^c$ 	& N			& NR			& PR			& CR	 \\ \hline
\endfirsthead

\multicolumn{7}{c}%
{{\bfseries \tablename\ \thetable{} -- continued from previous page}} \\

 				& 					& 				& \multicolumn{4}{c}{Our Results$^a$ }\\ 
Medication$^b$ 				& Dose 				& Frequency$^c$ 	& N			& NR			& PR			& CR	 \\ \hline
\endhead

\multicolumn{7}{r}{{\em Table cont'd on next page.}}\\ \hline
\endfoot
\hline\hline
\endlastfoot

Irinotecan$^d$				& 125 mg/m$^2$		& weekly,			& 320		& 1.9\%		& 80.9\%		& 17.2\% \\
and Cmab				& 400 mg/m$^2$		& load \&			& 			& 			&			& \\		
{}						& 250 mg/m$^2$		& weekly			& 			& 			&			& \\ 
 & & & & & \\
Irinotecan$^d$				& 125 mg/m$^2$		& weekly \&		& 320		& 14.4\%		& 67.5\%		 & 18.1\% \\
and Pmab					& 6 mg/kg			& q2w			& 			& 			&			& \\
 & & & & & \\
Irinotecan					& 125 mg/m$^2$		& weekly \&		& 320		& 3.4\%		& 84.4\%		& 12.2\%\\
and Cmab				& 400 mg/m$^2$		& load, day 4 \&	& 			& 			&			& \\		
{}						& 250 mg/m$^2$		& weekly			& 			& 			&			& \\ 
 & & & & & \\
Irinotecan					& 125 mg/m$^2$		& weekly, day4 \&	& 320		& 8.4\%		& 80.1\%		& 11.4\% \\
and Pmab					& 6 mg/kg			& q2w,			&			& 			&			& \\
 & & & & & \\
Irinotecan					& 350 mg/m$^2$  		& q3w			& 320		& 0\%		& 39.0\%		& 60.9\% \\	
and Cmab				& 500 mg/m$^2$		& q2w			&			& 			&			& \\ 
 & & & & & \\
Irinotecan					& 350 mg/m$^2$		& q3w \&			& 320		& 16.6\%		& 71.3\%		& 12.2\% \\
and Pmab					& 9 mg/kg			& q3w			&			& 			&			& \\ \hline

\label{tab:response_rates_my_tx}
\end{longtable}
\begin{flushleft}Response rates from clinical trial simulations for our experimental treatment schedules.\\
$^a$ N=number of patients; NR=no response; PR=partial response, CR=complete response.\\
$^b$ Pmab=panitumumab; Cmab=cetuximab\\
$^c$ q3w=every three weeks; q2w=every two weeks; load=loading dose.\\
$^d$ The standard treatments.
\end{flushleft}
\end{center}



\end{document}